\documentclass{aa}  
\usepackage{graphicx}
\usepackage{rotating,tabularx}

\usepackage{txfonts}
\begin{document}

   \title{The XXL Survey: XLIII. The quasar radio loudness dichotomy exposed via radio luminosity functions obtained by combining results from COSMOS and XXL-S X-ray selected quasars}

    \titlerunning{The origin of radio emission in X-ray selected quasars}


   \author{Lana Ceraj\inst{1,2}
	\thanks{\emph{lceraj@irb.hr}}
	\and V. {Smol\v{c}i\'{c}} \inst{1}
	\and I. Delvecchio \inst{3}	
	\and A. Butler \inst{4} 
	\and K. {Tisani\'{c}} \inst{1, 2}
	\and J. Delhaize \inst{5}
	\and C. Horellou \inst{6}
	\and J. Kartaltepe \inst{7}
    \and K. Kolokythas \inst{8}	
    \and S. Leslie \inst{9}
	\and S. Marchesi  \inst{10}
	\and M. Novak \inst{9}
	\and M. Pierre \inst{3}
	\and M. Plionis \inst{11}
	\and E. Vardoulaki \inst{12}
	\and G. Zamorani \inst{13}
          }

   \institute{
   University of Zagreb, Physics Department, Bijeni\v{c}ka cesta 32, 
	10002 Zagreb, Croatia.
	\and Ru\dj{}er Bo\v{s}kovi\'{c} Institute, Bijeni\v{c}ka cesta 54, 10002 Zagreb, Croatia.
	\and CEA, IRFU, DAp, AIM, Universit\'e Paris-Saclay, Universit\'e Paris Diderot, Sorbonne Paris Cit\'e, CNRS, F-91191 Gif-sur-Yvette, France.
    \and International Centre for Radio Astronomy Research (ICRAR), University of Western Australia, 35 Stirling Hwy, Crawley WA 6009, Australia.
	\and Department of Astronomy, University of Cape Town, Private Bag X3, Rondebosch 7701, South Africa.
	\and Chalmers University of Technology, Dept of Space, Earth and Environment, Onsala Space Observatory, SE-439 92 Onsala, Sweden.	
	\and School of Physics and Astronomy, Rochester Institute of Technology, Rochester, NY 14623, USA.
	\and Centre for Space Research, North-West University, Potchefstroom 2520, South Africa.
    \and Max-Planck-Institut f\"{u}r Astronomie, K\"{o}nigstuhl 17, D-69117 Heidelberg, Germany.
	\and Department of Physics and Astronomy, Clemson University, Clemson, SC 29634, USA.
	\and National Observatory of Athens, Lofos Nymfon, Thession, Athens 11810, Greece.
	\and Max-Planck-Institut f\"{u}r Radioastronomie, Auf dem H\"{u}gel 69, 53121 Bonn, Germany.	
	\and INAF - Osservatorio di Astrofisica e Scienza dello Spazio - Bologna, Via Piero Gobetti 93/3, I-40129 Bologna, Italy.}


 
\abstract{We studied a sample of 274 radio and X-ray selected quasars (XQSOs) detected in the COSMOS and XXL-S radio surveys at 3 GHz and 2.1 GHz, respectively. This sample was identified by adopting a conservative threshold in X-ray luminosity, $\mathrm{L_{X}\ [2-10\ keV] \geq 10^{44}\ erg\,s^{-1}}$, selecting only the most powerful quasars. 
Using available multiwavelength data, we examined various criteria for the selection of radio-loud (RL) and radio-quiet (RQ) XQSOs, finding that the number of RL/RQ XQSOs changes significantly depending on the chosen criterion. This discrepancy arises due to the different criteria tracing different physical processes and due to our sample being selected from flux-limited radio and X-ray surveys.
Another approach to study the origin of radio emission in XQSOs is via their radio luminosity function (RLF). We constructed the XQSO 1.4 GHz radio luminosity functions (RLFs) in six redshift bins at $\mathrm{0.5 \leq z \leq 3.75}$. The lower-1.4 GHz luminosity end shows a higher normalization than expected only from AGN contribution in all studied redshift bins. The found "bump" is mostly dominated by emission due to star-forming (SF) processes within the XQSO host galaxies. As expected, AGN-related radio emission dominates at the higher-luminosity end of RLF. 
The evolution of XQSO RLF was studied via combination of analytic forms from the literature to constrain the lower-luminosity "bump" and the higher-luminosity AGN part of the RLF. 
We defined two 1.4 GHz luminosity thresholds, $\mathrm{L_{th, SF}}$ and $\mathrm{L_{th, AGN}}$, below and above which more than $\mathrm{80\, \%}$ of sources contributing to the RLF are dominated by SF and AGN-related activity, respectively. These thresholds evolve with redshift, most likely due to the strong evolution of SFRs of the XQSO host galaxies.
Our results expose the dichotomy of the origin of radio emission: while the higher-luminosity end of the XQSO RLF is dominated by AGN activity, the lower-luminosity end is dominated by the star formation-related processes.}


   \maketitle


\section{Introduction}

Quasars are among the most powerful active galactic nuclei (AGN) and are usually very bright over most of the electromagnetic spectrum, from gamma to infrared frequencies. However, only $\mathrm{5-10\%}$ of quasars are associated with strong radio sources (e.g., \citealt{condon80}, \citealt{condon81}). This observation triggered a long lasting search for the physical origin of the quasars' radio emission, especially for type-1 (unobscured) quasars (e.g., \citealt{kellermann89}, \citealt{white07}, \citealt{balokovic12}, \citealt{condon13}). The most common approach is to separate the quasar population into so-called "radio-loud" (RL) and "radio-quiet" (RQ) categories with respect to some threshold (e.g., \citealt{ivezic02}, \citealt{cirasuolo03}, \citealt{pierce11}). There are, however, several definitions of radio loudness in the literature (\citealt{hao14}).

The most direct measure of radio loudness is the monochromatic radio luminosity. One of the first studies of radio emission in optically selected quasars focused on the distribution of 5 GHz radio luminosity, $\mathrm{L_{5\ GHz}}$ (\citealt{miller90}). \citeauthor{miller90} found a deficit of quasars with $\mathrm{10^{24}\ W\,Hz^{-1} \lesssim L_{5\ GHz}\lesssim 10^{25}\ W\,Hz^{-1}}$, with quite a few detections with $\mathrm{L_{5\ GHz}\gtrsim 10^{25}\ W\,Hz^{-1}}$ which made their radio luminosity distribution appear bimodal. They suggested that this bimodality is the evidence for the existence of two distinct populations of quasars above and below $\mathrm{L_{5\ GHz} \approxeq 10^{25}\ W\,Hz^{-1}}$.

In search for the existence of two physically distinct populations, most authors have defined radio loudness as the radio-to-optical flux density or luminosity ratios (e.g., \citealt{kellermann89}, \citealt{ivezic02}, \citealt{white07}, \citealt{balokovic12}). {Such a definition relates the two mechanisms associated with the accretion onto the supermassive black hole: (a) the optical luminosity (due to thermal emission from the accretion disk) and (b) the radio emission (due to synchrotron radiation from the core, jets or lobes). Bimodality in the distribution of the optical-to-radio emission would point to RL and RQ quasars having different sources of the dominant radiative output. Despite the same or similar definition of radio loudness, the results differ widely: some authors claim to have found a dichotomy (e.g., \citealt{ivezic02}, \citealt{white07}), while others see no evidence of it (e.g., \citealt{cirasuolo03}). As discussed by \citet{balokovic12}, the disagreement can be explained by selection biases not fully accounted for in the various analysis and low number statistics.}

Another approach to the problem of quasar radio loudness dichotomy is to study the quasar radio luminosity function (RLF; e.g., \citealt{kimball11}, \citealt{condon13}). In the local universe radio luminosity function is dominated by AGN-related radio emission above $\mathrm{L_{1.4\ GHz} \geq 10^{23}\ W\,Hz^{-1}}$, while radio emission from star-forming processes in galaxies starts to dominate below that threshold (e.g., \citealt{condon02}). \citet{condon13} studied the 1.4 GHz RLF of color-selected quasars from the Sloan Digital Sky Survey (SDSS) Data Release 7 Quasar catalog (\citealt{schneider10}) at redshifts $\mathrm{0.2<z<0.45}$ detected at 1.4 GHz within National Radio Astronomy Observatory (NRAO) Very Large Array (VLA) Sky Survey (NVSS; \citealt{condon98}). They argue that their RLF is most likely dominated by AGN-related radio emission due to the high radio luminosity ($\mathrm{L_{1.4\ GHz}\geq 10^{24.2}\ W\,Hz^{-1}}$) of the selected sources. To further study the origin of radio emission, they modelled the distribution of the peak flux densities of 179 color-selected quasars from the NVSS at redshifts $\mathrm{0.2<z<0.3}$ using flux densitites below the NVSS detection limit ($\mathrm{S_{1.4\ GHz}=2.5\ mJy}$; \citealt{condon98}). They argued that the observed distribution of peak fluxes for undetected sources should require the presence of a "bump" in the 1.4 GHz RLF centered at the luminosity $\mathrm{L_{1.4\ GHz}\approx 10^{22.7}\ W\,Hz^{-1}}$. This "bump", due to highly star-forming galaxies in Condon's model (see Fig. 6 in \citealt{condon13}), was confirmed by \citet{kimball11}, who examined the same sample of color-selected quasars from SDSS, individually detected at 6 GHz using the Expanded VLA (EVLA; \citealt{perley11}). They constructed the 6 GHz quasar RLF, finding the sources above and below $\mathrm{L_{6\ GHz} = 10^{23.5}\ W\,Hz^{-1}}$ to be dominated by AGN-related emission and star-forming processes, respectively. 

In this paper, we follow the approach of \citet{kimball11} and \citet{condon13} to study the origin of radio emission in radio and X-ray selected quasars (XQSOs) via RLFs. Our selection ($\mathrm{L_{X}[2-10\ keV]\geq 10^{44}\ erg\,s^{-1}}$) is aimed at tracing the sources exhibiting quasar activity in the COSMOS and XMM extragalactic survey south (XXL-S) fields. The datasets and the selection of our XQSO sample are described in Sect. \ref{sec:ds}. In Sect. \ref{sec:rld} we examine several definitions of radio loudness applied to our XQSOs. The derivation and evolution of the XQSO RLF are described in Sect. \ref{sec:xqsorlf}, and discussed in Sect. \ref{sec:discussion} in the context of other studies of the origin of radio emission in quasars. The main conclusions of this work are summarized in Sect. \ref{sec:sac}.

Throughout this paper, we assume a $\mathrm{\Lambda CDM}$ cosmology with $\mathrm{H_0 = 70\ km\,s^{-1}\,Mpc^{-1}}$, $\mathrm{\Omega_m = 0.3}$ and $\mathrm{\Omega_{\Lambda} = 0.7}$. We assume the radio spectrum can be described by a simple power-law form of the flux density $S_{\nu}\propto \nu ^{\alpha}$, where $\nu$ and $\alpha$ are the frequency and spectral index, respectively.

\section{Data and sample} \label{sec:ds}

In this section, we describe the datasets and selection of the sample of XQSOs. To construct a sample of quasars that span the wide range of 1.4 GHz radio luminosities needed to constrain the shape and evolution of the RLF, we combine the available radio datasets with the multiwavelength (X-ray to infrared) coverage in the COSMOS and XXL-S fields, as described in Sect. \ref{sec:data}. In Sect. \ref{sec:sample} we describe our quasar selection based on the hard band $\mathrm{[2-10\ keV]}$ X-ray luminosity and the properties of the sample.

\subsection{Radio data with multiwavelength counterparts} \label{sec:data}

\subsubsection{COSMOS field} \label{sec:vla3}

Radio data from the VLA-COSMOS 3 GHz Large Project are a result of $384$ hours of observations carried out with the Karl G. Jansky Very Large Array (VLA) at 3 GHz (10 cm). Observations were centered at the COSMOS field, covering $2.6 \, \mathrm{deg}^2$ with a 1$\sigma$ sensitivity of $2.3$ $\mathrm{\mu Jy / beam}$ and at an angular resolution of 0.75$''$. A total of 10830 sources were detected with signal-to-noise ratios (S/N) of five or more. The observations and the extraction of radio sources are described in detail by \citet{smolcic17a}. The COSMOS field had been previously observed at 1.4 GHz with 1$\sigma$ sensitivity of $\sim$10-15 $\mathrm{\mu Jy / beam}$ at an angular resolution of 1.5$''$ across $2\, \mathrm{deg}^2$ (\citealt{schinnerer07}, \citealt{schinnerer10}), providing flux densities for about $\sim$30$\%$ of the radio sources detected at 3 GHz. For these sources, the spectral index $\alpha$ was calculated as the slope between the measurements at 1.4 GHz and 3 GHz in the $\mathrm{log(S_{\nu})-log(\nu)}$ plane. The spectral indices of sources undetected at 1.4 GHz were set to be $\mathrm{-0.7}$ as this is the average spectral index found for the full 3 GHz population is the VLA-COSMOS 3 GHz Large Project (\citealt{smolcic17a}).

The sources detected at 3 GHz were cross-matched with the optical-NIR COSMOS2015 catalog (\citealt{laigle16}), yielding a sample of 7729 radio detected sources with multiwavelength counterparts over the unmasked 1.77 $\mathrm{deg}^2$ area. The process of assigning multiwavelength counterparts to radio detections is described in detail by \citet{smolcic17b}. 
{Reliable spectroscopic redshifts were available for $\sim$35$\%$ of those sources, while the rest ($\sim$65$\%$) have a photometric redshift estimate ($\mathrm{\sigma_{\Delta z/1+z}=0.021}$ at $\mathrm{3<z<6}$ and $\mathrm{\sigma_{\Delta z/1+z}<0.01}$ at $\mathrm{z<3}$) drawn from the COSMOS2015 catalog. }

As detailed in \citet{laigle16}, the COSMOS2015 catalog was also cross-matched with X-ray data from the Chandra COSMOS-Legacy Survey (\citealt{civano16}, \citealt{marchesi16}). For 906 ($\sim$12$\%$) radio detected sources with optical-NIR counterparts, absorption-corrected intrinsic X-ray luminosities are available in the soft [0.5-2 keV], hard [2-10 keV] and total [0.5-10 keV] bands. 

\subsubsection{XXL-S field} \label{sec:xxls}

The 2.1 GHz observations of the 25 $\mathrm{deg}^2$ XXL-S field were carried out with the Australia Telescope Compact Array (ATCA) within the 2.1 GHz ATCA XXL-S survey, as described in \citeauthor{butler18a} (\citeyear{butler18a}, hereafter XXL Paper XVIII). {The 1$\sigma$ level sensitivity achieved from the analysis of the radio mosaic is $\sim$41 $\mathrm{\mu}$Jy/beam at a resolution of $\sim4.8''$.} The number of sources detected with S/N of five or more is 6287. For a subsample of sources which were unresolved in the full 2 GHz ATCA bandwidth, \citeauthor{butler18b} (\citeyear{butler18b}, hereafter XXL Paper XXXI) constructed radio mosaics of three ATCA sub-bands (centered at 1417 MHz, 2100 MHz and 2783 MHz) to estimate radio spectral indices of these sources based on their sub-band flux densities. This procedure yielded estimates of spectral indices for 3827/6287 ($\sim$61$\%$) sources.

By cross-matching the radio data and the XXL-S catalog of optical counterparts (Fotopoulou et al., in prep.), \citeauthor{ciliegi18} (\citeyear{ciliegi18}, XXL Paper XXVI) produced a sample of 4770 sources with multiwavelength coverage. After removing 12 sources classified as stars, the sample of 4758 sources was used in Paper XXXI to study star-forming galaxies and AGN found in high-excitation and low-excitation radio galaxies. Out of these, $\sim$23$\%$ have reliable spectroscopic redshifts, while for the rest ($\sim$77$\%$) photometric redshift estimates ($\mathrm{\sigma_{\Delta z/1+z}=0.062}$) are available.

The observations of the XXL-S with the XMM-Newton X-ray telescope (\citealt{pierre16}, XXL Paper I) achieved depths of $\mathrm{6\times10^{-15}\ erg\, s^{-1}}$ and $\mathrm{2\times10^{-14}\ erg\, s^{-1}}$ in the $[\mathrm{0.5-2\ keV}]$ and $[\mathrm{2-10\ keV}]$ bands, respectively. The X-ray fluxes were added to the optical counterpart catalog, as described in Paper VI. These fluxes were used to estimate the hard [2-10 keV] band X-ray luminosity.

\subsection{X-ray and radio selected quasars} \label{sec:sample}

\subsubsection{Sample selection} \label{sec:selection}

From the catalogs of radio detected sources with multiwavelength counterparts described above, we selected quasars based on their hard X-ray luminosity. 

We applied a criterion of $\mathrm{L_X[2-10\ keV] \geq 10^{44}\ erg\,s^{-1}}$, which broadly selects quasars (e.g., review by \citealt{padovani17}). Throughout this paper we refer to these X-ray and radio selected quasars as XQSOs. 

\begin{figure}[!h]
\includegraphics[width=\linewidth]{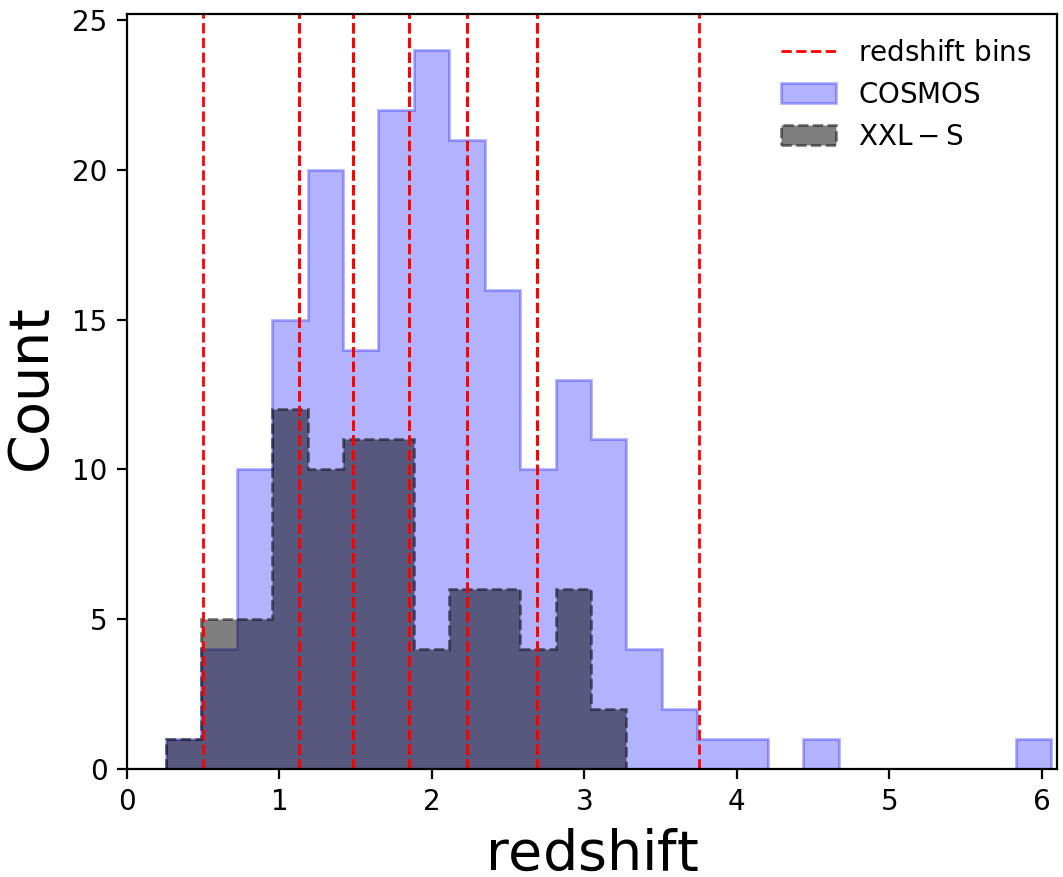}
\caption{Redshift distributions of COSMOS and XXL-S XQSOs are shown with blue and gray histograms, respectively. Red dashed lines show the edges of the redshift bins used in the radio luminosity function analysis in Sect. \ref{sec:rlf_evol}.}
\label{fig:z_dist}
\end{figure}

Our sample contains a total of 274 XQSOs, 191 from the COSMOS field and 83 from the XXL-S field. All COSMOS XQSOs were previously classified as moderate-to-high radiative luminosity AGN by \citet{smolcic17b}, while all XXL-S XQSOs were classified as high-excitation radio galaxies in Paper XXXI. For $\mathrm{60\%}$ of COSMOS XQSOs spectroscopic redshifts were available, while for the rest ($\mathrm{40\%}$) we use photometric redshifts ({with accuracy estimated to be $\mathrm{\sigma_{\Delta z/1+z, XQSO}=0.06}$}). All of the XXL-S XQSOs have available spectroscopic redshifts. The redshift distribution of XQSOs is shown in Fig. \ref{fig:z_dist}. With the selection based on the X-ray luminosities, our sample of COSMOS and XXL-S sources is complete out to $\mathrm{z\sim4}$ (see Fig. 7 by \citealt{marchesi16}) and $\mathrm{z\sim2.2}$ (see Fig. 7 in Paper XXXI), respectively.

\subsubsection{Infrared and B-band luminosities} \label{sec:irb}

For every source in the sample of XQSOs, the total infrared (IR; rest $\mathrm{8-1000\, \mu m}$) luminosity arising from star formation within the host galaxy was estimated from the multi-component (star-forming galaxy and AGN) best-fit template from broad-band SED-fitting decomposition, as presented by \citet{delvecchio17} and in Paper XXXI.
{For COSMOS sources, Herschel photometry from the COSMOS2015 catalog (\citealt{laigle16}) was used to constrain the far-infrared and submillimeter wavelength range of the source's SED. Out of 191 COSMOS XQSOs, 79 (53) are detected at $\mathrm{>3(>5)\, \sigma}$ levels in at least one Herschel band, which corresponds to $\mathrm{41\%}$ ($\mathrm{28\%}$) of the parent sample. For Herschel non-detections, $3\,\sigma$ upper limits were used in SED fitting.}
{On the other hand, the XXL-S field has no Herschel coverage and the SED fitting was performed in the UV-to-MIR range only. To estimate the reliability of the total IR luminosity of the XXL-S sources, the test described by \citeauthor{butler_phdt} (\citeyear{butler_phdt}; see Section 4.5.5. therein) was performed. Briefly, radio- and Herschel-detected sources in the COSMOS field were selected above the XXL-S sensitivity limit of the 2.1 GHz ATCA XXL-S radio survey. For these sources, after excluding FIR Herschel data, SED fitting was repeated. The resulting total IR luminosity estimates were compared to those derived including the FIR data, yielding a dispersion of $\sim0.3$ dex, with no systematic offset. Hence, the total IR luminosities obtained from SED fitting procedure are not strongly affected by the lack of Herschel detections.}

Using the same SED fitting procedure, the intrinsic 4058 \AA \ B-band AGN luminosity was estimated for all COSMOS sources and 81/83 XXL-S sources. Some of the B-band AGN luminosity estimates were calculated using SED templates in which the model of AGN component is not well constrained by the data. This is true for 23/191 of COSMOS and 14/81 of XXL-S sources. These values of B-band AGN luminosity can be considered lower limits.

\subsubsection{Radio spectral indices and luminosities} \label{sec:spectral_indices}

\begin{figure}[!h]
\includegraphics[width=\linewidth]{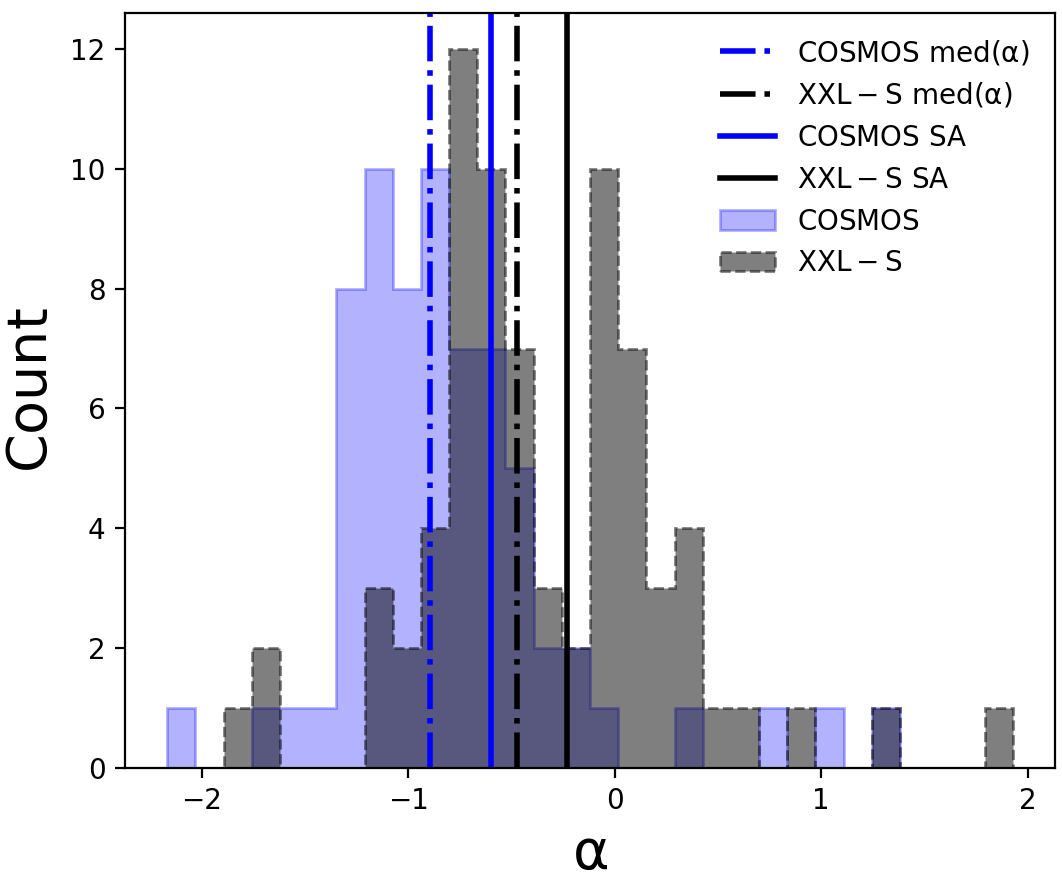}
\caption{Spectral indices of XQSOs in the COSMOS (blue) and XXL-S (gray) fields for the subsample of objects detected in more than one frequency. Dashed-dotted blue and black lines show the median spectral index of COSMOS and XXL-S sources, respectively. Solid blue and black lines show the value of spectral index obtained from the single-censored survival analysis (SA) of COSMOS and XXL-S sources, respectively.}
\label{fig:alpha_dist}
\end{figure}

The XQSO sample described above contains sources detected at 3 GHz (COSMOS) and 2.1 GHz (XXL-S). Spectral index estimates are available for 68 ($\sim$36$\%$) COSMOS and 75 ($\sim$90$\%$) XXL-S sources, as shown in Fig. \ref{fig:alpha_dist}. The median value and the standard deviation of the derived spectral indices are $-0.89 \pm 0.56$ and $-0.47 \pm 0.62$ for the COSMOS and XXL-S XQSOs, respectively. For the sources without a spectral index estimate, we apply the median values obtained from the single-censored survival analysis. For the sources within the COSMOS field, this method finds a value of ($-0.6 \pm 0.3$), while the estimated median spectral index value for the XXL-S sources is ($-0.23 \pm 0.12$). 

Spectral indices of COSMOS XQSOs are on average steeper than those of XXL-S sources due to flux limits of the radio surveys and the physical nature of the sources. As discussed by \citet{novak17}, in order for a 3 GHz source close to the 3 GHz detection limit to be detected at 1.4 GHz, it needs to have a very steep radio spectrum ($\mathrm{\alpha < -2}$). The spectral indices of the XXL-S sources were derived by splitting the observations at 2.1 GHz (bandwidth 2 GHz) into 3 sub-bands with bandwidths of 683 MHz centered at 1.417 GHz, 2.100 GHz and 2.783 GHz (Paper XXXI). Spectral indices derived in this way are less biased to the shape of the radio spectrum and a wide variety of spectral index values can be derived. As shown in Fig. \ref{fig:alpha_dist}, the distribution of XXL-S spectral indices appears to be double-peaked. The bin width was set to $\mathrm{3\times\sigma_{\alpha,XXL-S}}$, where $\mathrm{\sigma_{\alpha,XXL-S}=0.08}$ is the mean uncertainty on XXL-S spectral indices derived using sub-band analysis. We tested if there is dependence of spectral index on relevant physical parameters (such as radio flux densities and luminosities, X-ray luminosity and redshift) and found no statistically significant correlations. The bimodal shape of XXL-S XQSO spectral index distribution likely reflects the mix of core-dominated (i.e., those with flat radio spectral indices) and extended radio sources.

For XQSOs in the COSMOS field, both the median value of derived spectral indices and the estimate from the survival analysis are consistent with the spectral index $\mathrm{-0.7}$ usually quoted in the literature. This value is consistent with spectral indices found in literature in studies of radio spectra of both star-forming galaxies and AGN (e.g., \citealt{condon92}). On the other hand, roughly $\sim$50$\%$ of XXL-S XQSOs (and even more than $\mathrm{50\%}$ on the basis of the results from the survival analysis) have flat spectra ($\alpha>-0.5$) typical of core-dominated AGN radio emission. Hence, we expect the radio emission of XXL-S XQSOs to be dominated by an AGN contribution to the radio emission, while a mixture of two contributions (star formation and AGN activity) may contribute to the total observed radio emission of COSMOS XQSOs.

Observed flux densities $\mathrm{S_{\nu_{obs}}}$($\mathrm{W\,Hz^{-1}\,m^{-2}}$) and spectral indices are used to derive 1.4 GHz radio luminosity ($\mathrm{L_{1.4\ GHz}}$) defined as:
\begin{equation}\label{eq:14ghz_luminosity}
\mathrm{L_{1.4\ GHz} = \dfrac{4\pi D_{L}^{2}(z)}{(1+z)^{1+\alpha}}\bigg(\dfrac{1.4\ \mathrm{GHz}}{\nu_{obs}}\bigg)^{\alpha}S_{\nu_{obs}}},
\end{equation}
where $\mathrm{D_{L}\ [m]}$ is the luminosity distance, $\mathrm{z}$ is the redshift and $\mathrm{\nu_{obs}}$ is the observed frequency (3 GHz for COSMOS, 2.1 GHz for XXL-S). In the case of sources that were detected in multiple radio bands, the derived spectral index was used in the calculation. Otherwise, the median value obtained from the survival analysis was used (-0.6 for 123 COSMOS sources and -0.23 for 8 XXL-S sources). We used these 1.4 GHz luminosities to examine different radio loudness distributions and to construct XQSO RLFs.

\begin{figure}[!h]
\includegraphics[width=\linewidth]{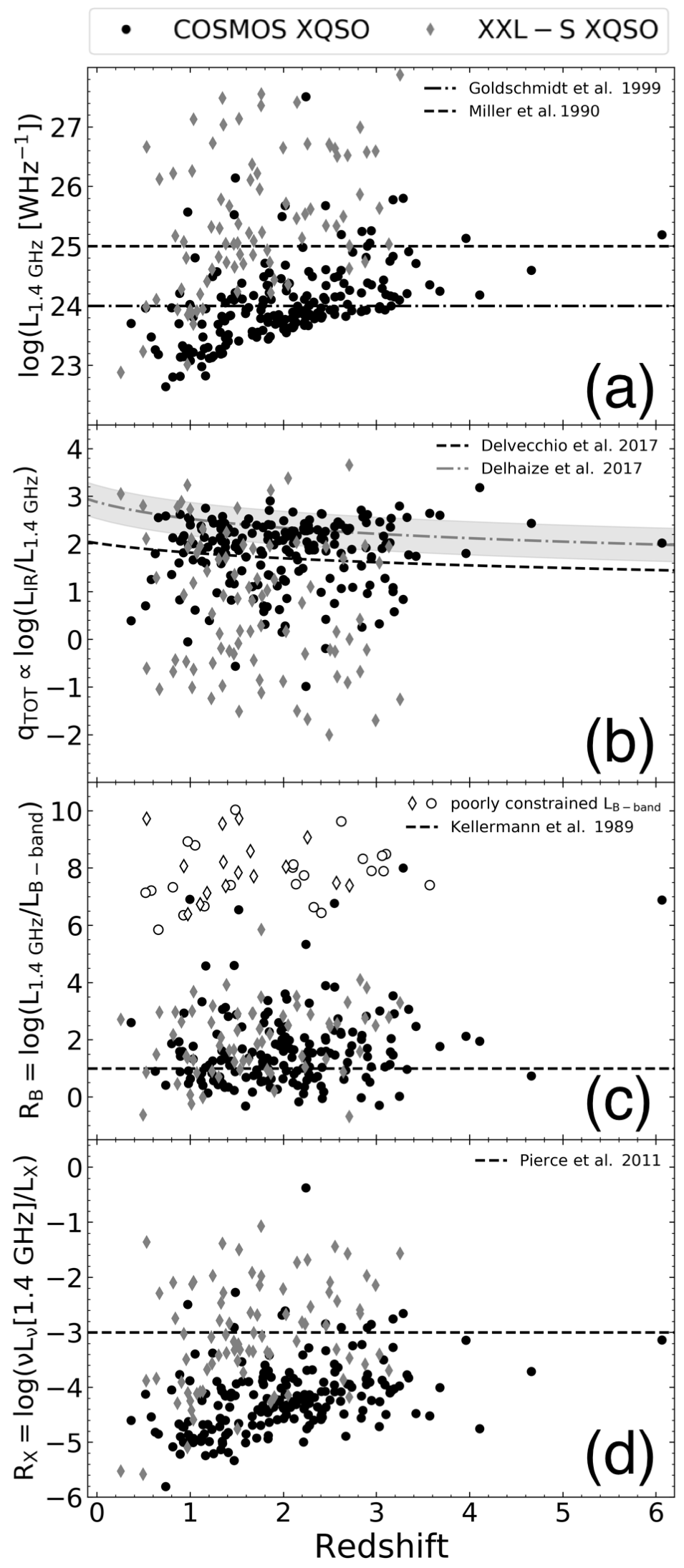}
\caption{{Radio loudness vs. redshift of XQSOs from COSMOS (circles) and XXL-S (diamonds). In different panels radio loudness is defined as: (a) 1.4 GHz radio luminosity, (b) IR-to-1.4 GHz luminosity ratio, $\mathrm{q_{TOT}}$, (c) 1.4 GHz radio-to-optical luminosity ratio, (d) 1.4 GHz radio-to-X-ray luminosity ratio. Different thresholds from the literature, used to select sources as RL, are shown as indicated in the legends. The gray dot-dashed line and area show the $\mathrm{q_{TIR}(z)\pm0.35}$ by \citet{delhaize17}.}}
\label{fig:distributions}
\end{figure}

\subsection{Possible resolution biases}

Throughout this paper, we use the IR luminosity due to star formation (constrained via SED fitting procedure which, when available, uses FIR Herschel observations) to estimate the contribution of star formation to the total radio emission of the XQSOs. 
An important point that needs to be considered when comparing infrared and radio luminosities expected to originate from the same region is the difference in the spatial resolution of the data from which they are derived. For the COSMOS XQSOs, which are detected both in radio (VLA) and infrared (Herschel), a difference in resolution could result in radio observations detecting smaller star-forming regions than the FIR observations. To test this, we checked the fraction of resolved sources within the radio-excess and star-forming subpopulations, as defined using the correlation by \citet{delvecchio17} as the threshold between the two classes. We found that radio-excess XQSOs seem to be systematically more resolved than star-forming ones, probably due to the presence of the large scale AGN-related structures. 
The physical size at the mean redshift of both AGN- and SF- dominated subsets ($\mathrm{z\sim2}$) is $\mathrm{8.4\ kpc}$ per arcsec. The VLA beam size ($\mathrm{0.75\, "}$) at $\mathrm{z=2}$ is $\mathrm{6.4\ kpc}$, which means that it is large enough to enclose a typical star-forming galaxy at that redshift (e.g., \citealt{vdw14}). Hence, we do not expect radio observations to miss flux related to star formation (see also \citealt{delhaize17}).
The FIR Herschel fluxes have been extracted in a way to recover the entire FIR flux of the galaxy (for more details see \citealt{laigle16}).
We hence conclude that the difference between the resolutions of radio and infrared observations should not significantly affect the flux comparison between the two bands.

\section{Radio loudness} \label{sec:rld}

In this section, we compare four different definitions of radio loudness and criteria used to select RL and RQ AGN. Each of these criteria relies on a different spectral window into AGN activity: (a) radio emission from the cores, jets or lobes; (b) excess of radio emission relative to that expected only from star-forming processes within the host galaxies; (c) AGN-related optical emission from the accretion disk; and (d) X-ray emission from the hot corona.

\begin{enumerate}[(a)]
\item \textbf{1.4 GHz radio luminosity ($\mathrm{L_{1.4\ GHz}}$)} 

The simplest criterion is based solely on the radio luminosity of the source. \citet{miller90} and \citet{gold99} analyzed the distribution of the 5 GHz radio luminosity of optically-selected quasars. To test the thresholds they applied to radio luminosity, we convert the 5 GHz into 1.4 GHz luminosity thresholds with the cosmology used in this work. The RL quasars are then defined as those with 1.4 GHz luminosity above $\mathrm{L_{1.4\ GHz} \approx 10^{24}\ W\,Hz^{-1}}$ (\citealt{gold99}) and $\mathrm{L_{1.4\ GHz} \approx  10^{25}\ W\,Hz^{-1}}$ (\citealt{miller90}).

\item \textbf{IR-to-radio luminosity ratio ($\mathrm{q_{TOT}}$)} 

The logarithmic value of the ratio of the IR-to-1.4 GHz luminosity, $\mathrm{q_{TOT}\propto log(L_{IR}/L_{1.4\ GHz})}$, can be used to select sources which display an excess of radio emission due to AGN activity in relation to that expected just from star-forming processes in the host galaxy (e.g., \citealt{delvecchio17}, \citealt{smolcic17c}, \citealt{ceraj18}). For example, \citet{delvecchio17} derived a redshift-dependent radio excess threshold ($\mathrm{q_{REX}= 24-21.984\times(1+z)^{0.013}}$) selecting sources in which there is negligible ($\sim$0.15$\%$) contamination of radio emission due to star formation. We select sources that display significant radio excess ($\mathrm{q_{TOT}\leq q_{REX}}$) as RL. 
{However, it is important to note that the $\mathrm{q_{TOT}}$ is not a direct estimate of radio loudness as it relates IR luminosity from star formation and radio luminosity, which is a combination of both star-forming processes and AGN activity, unlike standard radio loudness definitions which compare AGN-related emissions only.}

\item \textbf{Radio-to-optical luminosity ratio ($\mathrm{R_{B}}$)} 

The most widely used criterion to separate RL and RQ quasars is based on the radio-to-optical flux density or luminosity ratio (e.g., \citealt{kellermann89}, \citealt{ivezic02}, \citealt{balokovic12}). RL quasars are defined as those having radio emission a factor of $\sim$10 stronger than the optical emission, i.e. $\mathrm{R_{B}=log(L_{1.4\ GHz}/L_{B-band})=1}$, where $R_{B}$ is the logarithmic ratio of radio-to-B-band luminosity. In Fig. \ref{fig:distributions}, empty symbols show the upper limits of $\mathrm{R_{B}}$ calculated using poorly constrained B-band AGN luminosities (see Sect. \ref{sec:irb}).

\item \textbf{Radio-to-X-ray luminosity ratio ($\mathrm{R_{X}}$)} 

Radio loudness defined as the logarithm of radio-to-$\mathrm{[2-10\ keV]}$ X-ray luminosity ratio, $\mathrm{R_{X}=log(\nu\, L_{\nu} (1.4\ GHz)/L_{X}[2-10\ keV])}$, has proven to be successful in selecting RL AGN even in cases of heavily obscured nuclei (e.g., \citealt{terashima03}). Following \citet{pierce11}, we adopt the threshold $\mathrm{R_{X}\approx-3}$ to separate RL ($\mathrm{R_X>-3}$) and RQ ($\mathrm{R_X<-3}$) quasars.
\end{enumerate}

\begin{figure*}[!h]
\includegraphics[width=\linewidth]{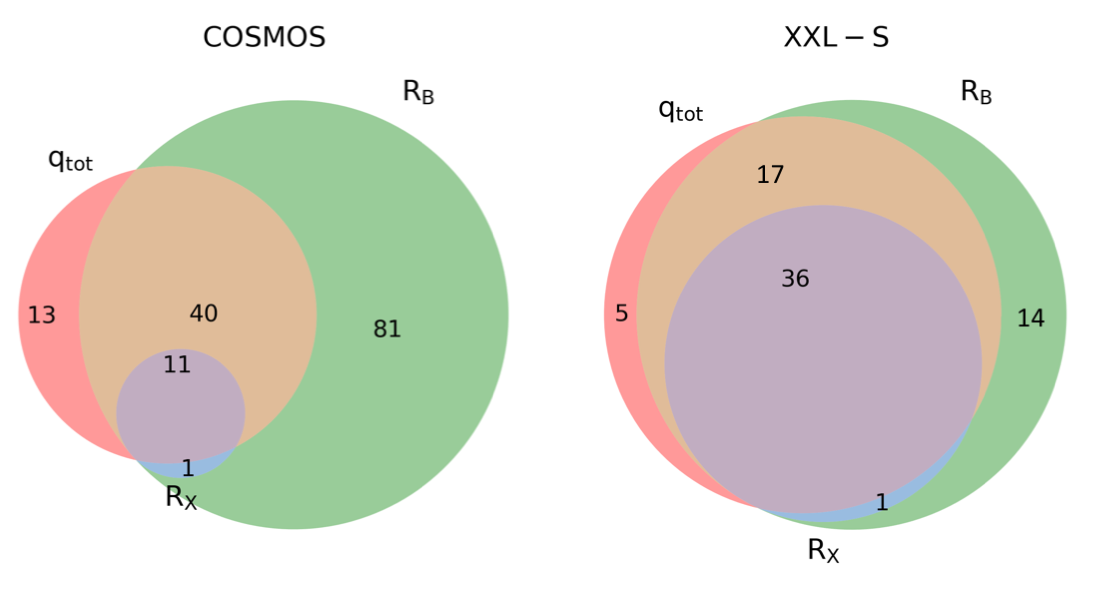}
\caption{{Venn diagram of COSMOS (left) and XXL-S (right) XQSOs classified as radio-loud based on three criteria: i) radio-excess threshold by \citet{delvecchio17} (pink circles), ii) $\mathrm{R_{B}}$ threshold by \citet{kellermann89} (green cricles) and iii) $\mathrm{R_{X}}$ threshold by \citet{terashima03} (blue circles).}}
\label{fig:venn}
\end{figure*}

The radio loudness distributions described above are shown in Fig. \ref{fig:distributions} as a function of redshift. {The number of RL and RQ XQSOs, as defined by different criteria, are listed in Table \ref{tab:radio_loudness}, and numbers of sources classified as RL using criteria (b)-(d) are shown using Venn diagrams in Fig. \ref{fig:venn}.
It is important to note that, for the same sample of sources, these criteria lead to a selection of different numbers of RL XQSOs ranging from $\sim$18$\%$ to $\sim$73$\%$. The origin of this discrepancy is the difference in definitions (a) and (b), which are related to different physical mechanisms than (c) and (d). The radio loudness defined via either monochromatic radio luminosity ($\mathrm{L_{1.4\ GHz}}$) or the IR-to-radio luminosity ratio ($\mathrm{q_{TOT}}$) traces the excess of radio emission, with RL sources being those with high levels of radio emission and those in which there is more radio emission detected than expected from the star formation within the host galaxy, respectively. On the other hand, radio loudness defined as $\mathrm{R_{B}}$ and $\mathrm{R_{X}}$ compares radiative (traced via B-band or X-ray luminosities) and kinetic (traced via radio luminosity) AGN emission. By applying (a)-(b) definitions of radio loudness, we found that COSMOS XQSOs are predominantly RQ, while XXL-S XQSOs are mostly RL (see Table \ref{tab:radio_loudness}). This is a result of different flux density limits of COSMOS and XXL-S radio surveys.}

\begin{figure}[!h]
\includegraphics[width=\linewidth]{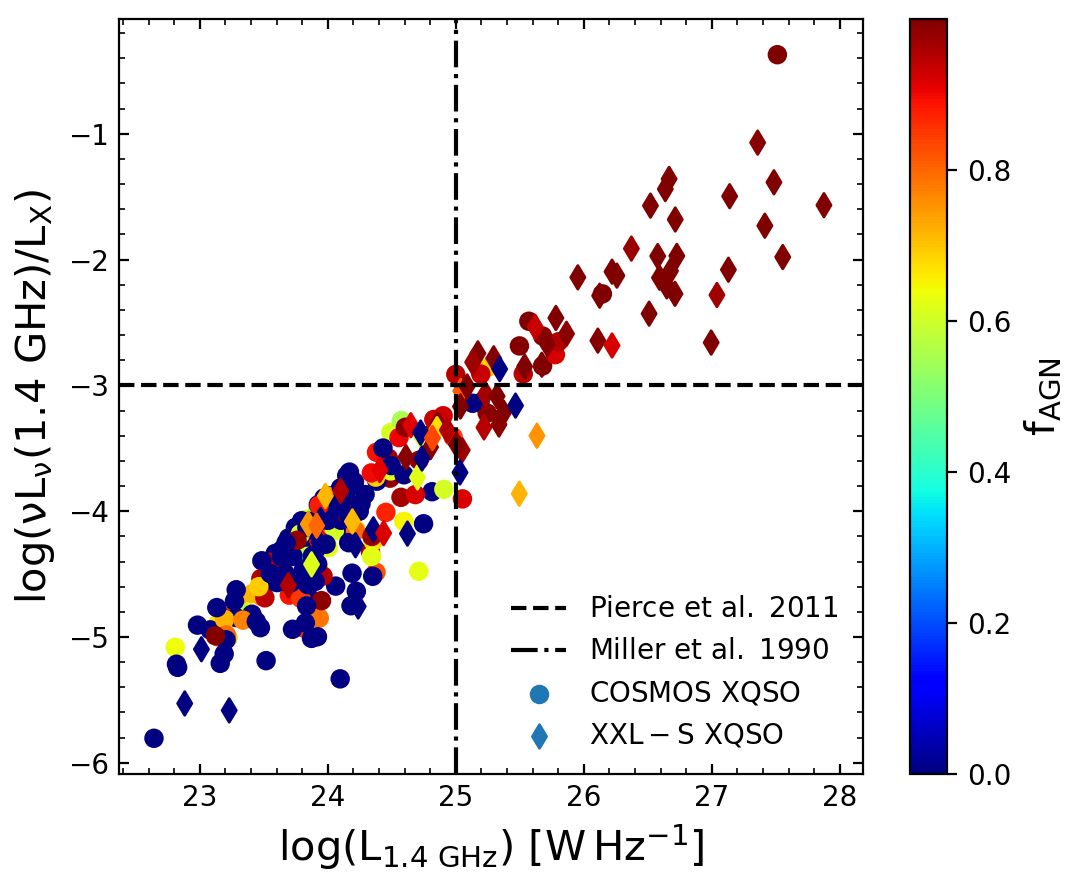}
\caption{$\mathrm{R_{X}}$ vs. 1.4 GHz luminosity color-coded by the AGN fraction ($\mathrm{f_{AGN}}$) of COSMOS (circles) and XXL-S (diamonds) XQSOs. Radio loudness threshold in radio luminosity as defined by \citet{miller90} is shown with the black dash-dotted vertical line, while the radio-to-X-ray radio loudness threshold from \citet{pierce11} is shown with the black dashed horizontal line. It can be seen that both RL thresholds miss a significant fraction of sources where the AGN dominates the radio emission.}
\label{fig:Lx_Lr}
\end{figure}

\begin{table*} [ht]
\begin{center}
\caption{Number of XQSOs defined as RL or RQ based on different criteria of radio loudness. {In the number count of XXL-S sources using a B-band based radio loudness estimate, two sources without B-band luminosity estimate were excluded, while those with poorly constrained B-band luminosities were counted as being radio-loud.}}
\renewcommand{\arraystretch}{1.5}
\quad
\renewcommand{\arraystretch}{1.5}
\begin{tabular}[t]{c c c c c c c c}
\hline
& \multicolumn{3}{c}{$\text{COSMOS}$} 
& \multicolumn{3}{c}{$\text{XXL-S}$} 
& \text{Total}\\ 
Threshold & \text{N(RL)} & \text{N(RQ)} & \text{$\mathrm{f_{RL}}$($\%$)} & \text{N(RL)} & \text{N(RQ)} & \text{$\mathrm{f_{RL}}$($\%$)} & \text{$\mathrm{f_{RL}}$($\%$)} \\
\hline  
$\mathrm{L_{1.4\ GHz} = 10^{24}\ WHz^{-1}}$ & 86 & 105 & 45.0 & 72 & 11 & 86.7 & 57.7\\
$\mathrm{L_{1.4\ GHz} = 10^{25}\ WHz^{-1}}$ & 16 & 175 & 8.4 & 52 & 31 & 62.7 & 24.8\\
$\mathrm{q_{REX} = 24 - 21.984(1+z)^{0.013}}$ & 64 & 127 & 33.5 & 58 & 25 & 69.9 & 44.5\\
$\mathrm{R_B = 1}$ & 133 & 58 & 69.6 & 66 & 15 & 81.5 & 73.2\\
$\mathrm{R_x = -3}$ & 12 & 179 & 6.3 & 37 & 46 & 44.6 & 17.9 \\

\hline
\end{tabular}
\label{tab:radio_loudness}
\end{center}
\end{table*}

As described in Sect. \ref{sec:spectral_indices}, the spectral indices of the COSMOS XQSOs are on average steeper than those of XXL-S XQSOs, which suggests a different origin of radio emission in these sources. The steep spectral indices of COSMOS XQSOs, which are mostly RQ, can originate from both star formation within the host galaxy and AGN. The logarithm of IR-to-1.4 GHz radio luminosity ratio, $\mathrm{q_{TOT}}$, can be used to quantify radio excess, as was previously done by \citet{ceraj18} for all COSMOS sources detected at 3 GHz that have multiwavelength coverage. By quantifying the radio excess due to an AGN contribution to the total radio emission, relative to that expected from the star forming-related IR luminosities, \citeauthor{ceraj18} estimated the AGN fractions, $\mathrm{f_{AGN}}$, defined as the AGN-related contribution to the total detected radio emission. Following the same approach to estimate $\mathrm{f_{AGN}}$ of XQSOs, we use the infrared-radio correlation derived by \citeauthor{delhaize17} (\citeyear{delhaize17}; shown in Fig. \ref{fig:distributions}(b)) derived from a sample of $\sim$9500 star-forming galaxies (SFGs) in the COSMOS field. They found a redshift-dependent infrared-to-1.4 GHz radio luminosity ratio, $\mathrm{q_{TIR}(z)=2.88\times(1+z)^{-0.19}}$, with a spread of 0.35 dex. Under the assumption that the host galaxies of XQSOs are regular SFGs (as supported by results by, e.g., \citealt{stanley15}), we can use $\mathrm{q_{TIR}(z)}$ to derive AGN fractions in XQSOs defined as $\mathrm{f_{AGN} = 1 - 10^{q_{TOT}-q_{TIR}}}$. 

Fig. \ref{fig:Lx_Lr} shows the logarithm of radio-to-$\mathrm{[2-10\ keV]}$ X-ray luminosity ratio, $\mathrm{R_{X}}$, as a function of $\mathrm{L_{1.4\ GHz}}$, color-coded by the AGN fraction from both COSMOS and XXL-S sources. {The radio loudness defined solely as the threshold in 1.4 GHz luminosity (\citealt{miller90}) mostly selects as RL sources which are dominated by the AGN-related radio emission. However, there are a few outliers ($\mathrm{5/68}$) in which the AGN fraction is $\mathrm{f_{AGN}<0.5}$. For all sources with $\mathrm{R_X}$ above the threshold $\mathrm{R_X=-3}$ radio emission is dominated by AGN activity ($\mathrm{f_{AGN} \geq 0.5}$), with the exception of one XXL-S source. 
Below the thresholds adopted for both criteria, there are roughly $\sim$50$\% \ (102/206)$ of sources which are dominated by AGN-related radio emission ($\mathrm{f_{AGN} \geq 0.5}$). This shows that applying a simple radio loudness threshold yields only subsamples of AGN-dominated sources in the radio.} In order to study the origin of radio emission in XQSOs, another approach is needed.

The radio luminosity function (RLF) has proven to be a highly useful tool for the study of the origin of radio emission in optically-selected quasars (\citealt{kimball11}, \citealt{condon13}). {In the next sections we follow the \citeauthor{kimball11} approach in the study of the origin of the detected radio emission by constructing RLFs for all radio-detected XQSOs at $\mathrm{0.5<z<3.75}$. For these sources we examine the cosmic evolution of the star-forming and AGN contributions to the radio emission via an analysis of the shape of their RLF.}

\begin{figure*}[ht]
  \includegraphics[width=\linewidth]{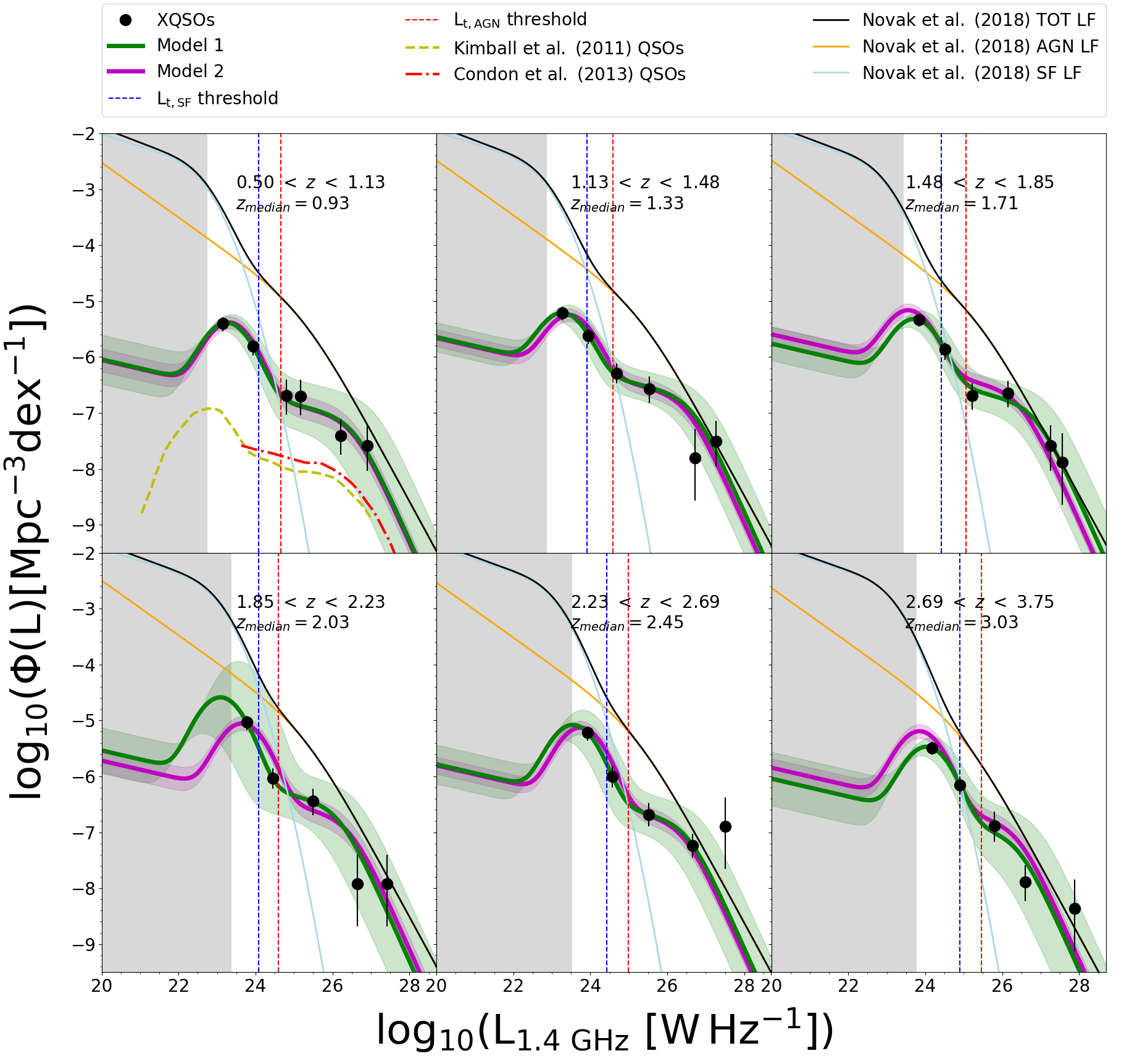}
  \caption{Radio luminosity function at 1.4 GHz of the radio and X-ray selected quasars at redshifts $0.50<z<3.75$. Solid green and magenta lines show the results of testing Model 1 and Model 2, respectively, with shaded areas showing 1$\mathrm{\sigma}$ confidence range. Thin solid light blue, orange, and black lines show the SFG, AGN and the total (SFG and AGN) RLFs by \citet{novak18}. Thin blue and red vertical dashed lines show the threshold luminosities below and above which more than 80$\%$ of sources contributing to the RLF are dominated by SF and AGN-related processes, respectively. {The gray shaded area shows the region unconstrained by our data where the RLF is just an extrapolation. Our results imply that the radio emission in a majority of lower radio luminosity XQSOs is powered by star formation.}}
  \label{fig:xqso_rlf}\vspace{1cm}
\end{figure*}

\section{Radio luminosity functions and their cosmic evolution} \label{sec:xqsorlf}

In this section, we describe the derivation of the radio luminosity function (RLF) of 267 XQSOs in six redshift bins (see Fig. \ref{fig:z_dist}) with roughly the same number of sources ($\sim$45) in each bin over the range $\mathrm{0.50 \leq z \leq 3.75}$ (Sect. \ref{sec:lf_derivation}). We then describe the choice of the local radio luminosity functions (Sect. \ref{sec:loc_rlf}) used to constrain the evolution of the XQSO radio luminosity functions (Sect. \ref{sec:rlf_evol}).

\subsection{Derivation} \label{sec:lf_derivation}

To derive the radio luminosity functions of XQSOs, we followed the $\mathrm{V_{max}}$ procedure to calculate the maximum observable volume for each source in our sample (\citealt{schmidt68}):
\begin{equation}\label{eq:vmax}
\mathrm{V_{max,i} = \int_{z_{min}}^{z_{max}} \mathcal{C}(S_{\nu})\frac{dV}{dz}dz},
\end{equation}
where $\mathrm{V_{max,i}}$ is the maximum observable volume of the i-th source within the considered redshift bin, $\mathrm{z_{min}}$ is the minimum redshift of the redshift bin and $\mathrm{z_{max}}$ is the maximum redshift at which a source of given radio luminosity could still be detected in our surveys. The factor which takes into account the size of the field and corrects for the incompleteness of the radio catalogs, $\mathrm{\mathcal{C}}$, is defined as:
\begin{equation}\label{eq:correction}
\mathrm{\mathcal{C}(S_{\nu}) = \dfrac{A_{obs}}{A_{sky}} \times \mathcal{C}_{comp}(S_{\nu})},
\end{equation}
where $\mathrm{A_{obs}}$ is the unmasked area of the fields considered (1.77 $\mathrm{deg^{2}}$ in COSMOS, 23.32 $\mathrm{deg^{2}}$ in XXL-S) and $\mathrm{A_{sky} \approx 41253\ deg^2}$ is the area of the celestial sphere. The completeness of the radio catalogs, $\mathrm{\mathcal{C}_{comp}(S_{\nu})}$, is taken from \citeauthor{smolcic17a} (\citeyear{smolcic17a}; see their Table 2) and Paper XVIII (see their Fig. 11).

Radio luminosity functions, $\mathrm{\Phi(L)}$, were calculated as the number of sources per co-moving volume per interval of logarithm of luminosity as:
\begin{equation}\label{eq:rlf}
\mathrm{\Phi(L)=\dfrac{1}{\Delta log(L)}\sum_{i=1}^{N}\dfrac{1}{V_{max,i}}},
\end{equation}
where $\mathrm{V_{max,i}}$ is the maximum observable volume and $\mathrm{\Delta log(L)}$ is the logarithmic luminosity bin width, taken to be 0.8 dex here. Errors on $\mathrm{\Phi(L)}$ are calculated following \citet{marshall85}:
\begin{equation}\label{eq:rlf_errors}
\mathrm{\sigma_{\Phi}(L)=\dfrac{1}{\Delta log(L)}\sqrt{\sum_{i=1}^{N}\dfrac{1}{V_{max,i}^2}}}.
\end{equation}
If a luminosity bin contains fewer than 10 sources, we corrected the errors on $\mathrm{\Phi(L)}$ for small number statistics using Poisson uncertainty estimates by \citet{gehrels86}. The XQSO RLFs, constructed using the combined sample of COSMOS and XXL-S sources, are shown in Fig. \ref{fig:xqso_rlf} and listed in Table \ref{tab:xqso_rlf}. 
The procedure is consistent with that of \citet{novak17}, \citet{smolcic17c}, \citet{ceraj18} and \citet{novak18}. 

In our analysis, we do not correct our RLFs to account for the fact that the X-ray data used to select XQSOs in the XXL-S field are incomplete at $\mathrm{z>2.2}$ (see Sect. \ref{sec:selection}). The last two redshift bins are affected by this incompleteness and should be taken with caution. We note that our results remain valid even if we do not consider the last two redshift bins.

\begin{table*}
\begin{center}
\caption{Radio luminosity functions of the radio and X-ray selected quasars. $\mathrm{log(L_{1.4\ GHz})}$ is the logarithm of median value of $\mathrm{L_{1.4\ GHz}}$ of all sources within the luminosity bin and N is the number of sources per luminosity bin.}
\renewcommand{\arraystretch}{1.5}
\quad
\renewcommand{\arraystretch}{1.5}
\begin{tabular}[t]{c c c c c}
\hline
\hline
Redshift
& $\log\left(\dfrac{L_{1.4\,\text{GHz}}}{\text{W}\,\text{Hz}^{-1}}\right)$
& $\log\left(\dfrac{\Phi}{\text{Mpc}^{-3}\,\text{dex}^{-1}}\right)$ & $\text{N}$\\
\hline  
$0.50<z<1.13$ & 23.16 &  -5.39 $^{+0.11}_{-0.15}$ & 16 \\
$Med(z) = 0.93$ & 23.93 & -5.81 $^{+0.12}_{-0.17}$ & 17 \\
& 24.80 & -6.70 $^{+0.30}_{-0.34}$& 3 \\
& 25.17 & -6.70 $^{+0.30}_{-0.34}$& 3 \\
& 26.22 & -7.41 $^{+0.30}_{-0.34}$& 3 \\
& 26.90 & -7.58 $^{+0.37}_{-0.45}$& 2 \\
\hline
$1.13<z<1.48$ & 23.28 &  -5.21 $^{+0.10}_{-0.14}$ & 16 \\
$Med(z) = 1.33$ & 23.95 & -5.62 $^{+0.11}_{-0.15}$ & 13 \\
& 24.69 & -6.29 $^{+0.17}_{-0.18}$& 8 \\
& 25.53 & -6.57 $^{+0.22}_{-0.25}$& 5 \\
& 26.73 & -7.81 $^{+0.52}_{-0.76}$& 1 \\
& 27.26 & -7.51 $^{+0.37}_{-0.45}$& 2 \\
\hline
$1.48<z<1.85$ & 23.84 &  -5.34 $^{+0.09}_{-0.11}$ & 20 \\
$Med(z) = 1.71$ & 24.52 & -5.87 $^{+0.13}_{-0.19}$ & 12 \\
& 25.22 & -6.69 $^{+0.22}_{-0.25}$& 5 \\
& 26.15 & -6.65 $^{+0.22}_{-0.25}$& 5 \\
& 27.25 & -7.59 $^{+0.37}_{-0.45}$& 2 \\
& 27.55 & -7.89 $^{+0.52}_{-0.76}$& 1 \\
\hline
$1.85<z<2.23$ & 23.77 &  -5.03 $^{+0.11}_{-0.14}$ & 30 \\
$Med(z) = 2.03$ & 24.45 & -6.03 $^{+0.17}_{-0.18}$ & 8 \\
& 25.50 & -6.45 $^{+0.22}_{-0.25}$& 5 \\
& 26.65 & -7.92 $^{+0.52}_{-0.76}$& 1 \\
& 27.41 & -7.92 $^{+0.52}_{-0.76}$& 1 \\
\hline
$2.23<z<2.69$ & 23.93 &  -5.22 $^{+0.11}_{-0.14}$ & 25 \\
$Med(z) = 2.45$ & 24.58 & -6.00 $^{+0.19}_{-0.20}$ & 7 \\
& 25.52 & -6.68 $^{+0.20}_{-0.22}$& 6 \\
& 26.66 & -7.24 $^{+0.20}_{-0.22}$& 6 \\
& 27.51 & -6.89 $^{+0.52}_{-0.76}$& 1 \\
\hline
$2.69<z<3.75$ & 24.18 &  -5.49 $^{+0.10}_{-0.13}$ & 22 \\
$Med(z) = 3.03$ & 24.91 & -6.15 $^{+0.11}_{-0.15}$ & 13 \\
& 25.79 & -6.89 $^{+0.25}_{-0.28}$& 4 \\
& 26.59 & -7.88 $^{+0.30}_{-0.34}$& 3 \\
& 27.88 & -8.36 $^{+0.52}_{-0.76}$& 1 \\

\hline
\end{tabular}
\label{tab:xqso_rlf}
\end{center}
\end{table*}

\subsection{Local luminosity function} \label{sec:loc_rlf}

In order to constrain the XQSO RLF, we need to adopt an analytic representation of the local RLF from the literature. Neither COSMOS nor XXL-S sample large enough volumes to detect a large number of local XQSOs in the radio band. To constrain the evolution of the XQSO RLF over the studied redshifts, we modify the local RLFs from the work by \citet{kimball11} and \citet{condon13}.

As found by \citet{kimball11}, the local ($0.2<z<0.3$) RLF of optically selected QSOs is a superposition of contributions from star formation (SF) within the host galaxy (dominant at the low-luminosity end) and AGN-related radio emission (dominant at the high luminosity end). The radio emission from the host galaxies produces a "bump" at the lower-luminosity end of the RLF which can be represented with a parabolic function:
\begin{equation}\label{eq:rlf_sf}
\mathrm{\log\Phi_{SF,0}(L)=\log\Phi_{SF}^{*}-(\log L-\log L^*)^2},
\end{equation}
where $\mathrm{\Phi_{SF}^{*} = 1.59\times10^{-7}\ Mpc^{-3} dex^{-1}}$ and $\mathrm{L^{*}_{SF}=10^{22.84}\ W\,Hz^{-1}}$ are the normalization and luminosity position of the vertex of parabola, respectively, converted to the units used in this paper. 

To constrain the high luminosity end ($\mathrm{L_{1.4\ GHz} >10^{24}\ W\,Hz^{-1}}$) of the RLF, we modified the analytic form of the RLF as described by \citet{condon13}. \citeauthor{condon13} examined the RLF of the color-selected quasars from the SDSS at $\mathrm{0.2<z<0.45}$ using the NVSS 1.4 GHz radio data. They found that their data constrain only the high luminosity end of the RLF where the radio emission is dominated by AGN activity. They fitted their data with an analytic function which is a combination of a power law function below $\mathrm{L_{1.4\ GHz}=10^{25.8}\ W\,Hz^{-1}}$ and a quadratic function above it. Most of the analytic forms of RLFs in literature are some variation of the double power law function (e.g., \citealt{condon98}, \citealt{sadler02}, \citealt{ms07}, \citealt{pracy16}). To be consistent with the literature RLF, we adopted a double power law function. We constrained the parameters using the \citet{condon13} results for quasars in SDSS at $\mathrm{0.2<z<0.45}$ (de-evolved to $\mathrm{z=0.25}$). This double power law function is:
\begin{equation}\label{eq:rlf_agn}
\mathrm{\Phi_{AGN,0} (L)  = \dfrac{\Phi^{*}_{AGN}}{\left(L/L^{*}_{AGN}\right)^{\alpha} + \left(L/L^{*}_{AGN}\right)^{\beta}}},
\end{equation}
where $\mathrm{\Phi^{*}_{AGN}=4.52\times10^{-9}\ Mpc^{-3}dex^{-1}}$ is the normalization and $\mathrm{L_{AGN}^{*}=4.74\times10^{26}\ WHz^{-1}}$ is the knee of the luminosity function. The faint and bright end slopes of the RLF are $\mathrm{\alpha = 0.16}$ and $\mathrm{\beta = 1.54}$, respectively.

\subsection{Cosmic evolution of XQSOs} \label{sec:rlf_evol}

To test the redshift evolution of the RLF, we combined the analytic functions $\mathrm{\Phi_{SF,0}}$ and $\mathrm{\Phi_{AGN,0}}$ and assumed that both the SF and AGN components can evolve both in density and luminosity. The general analytic form of the function describing such evolution is:
\begin{equation}\label{eq:rlf_tot_evol}
\begin{aligned}
\mathrm{\Phi_{TOT}(L,z) = \ (1+z)^{\alpha_{D,SF}}\,\Phi_{SF,0} \left[ \dfrac{L}{(1+z)^{\alpha_{L,SF}}}\right]} \\
\mathrm{\, \, \, \, \,  +\, (1+z)^{\alpha_{D,AGN}}\Phi_{AGN,0}\left[ \dfrac{L}{(1+z)^{\alpha_{L,AGN}}} \right]},
\end{aligned}
\end{equation}
where $\mathrm{\alpha_{D,SF}}$, $\mathrm{\alpha_{L,SF}}$, $\mathrm{\alpha_{D,AGN}}$ and $\mathrm{\alpha_{L,AGN}}$ are parameters of the evolution of the SF and AGN components, respectively. 
The standard approach used in the literature (e.g., \citealt{pracy16}, \citealt{smolcic17c}, \citealt{ceraj18}) is to assume the simplest models of evolution: pure density ($\mathrm{\alpha_{L,SF},\alpha_{L,AGN} =0}$) or pure luminosity evolution ($\mathrm{\alpha_{D,SF},\alpha_{D,AGN} =0}$). However, due to significant variation in density and luminosity between different redshift bins, we found that such simplified models do not describe our data. In the following analysis, we use $\mathrm{\chi ^{2}}$ minimization to test models in which both density and luminosity of both SF and AGN components change with redshift. An outlier at $\mathrm{2.23<z<2.69}$ with $\mathrm{L_{1.4\ GHz}>10^{27}\ WHz^{-1}}$ was excluded from the following analysis.

\begin{figure}[!htbp]
  \includegraphics[width=\linewidth]{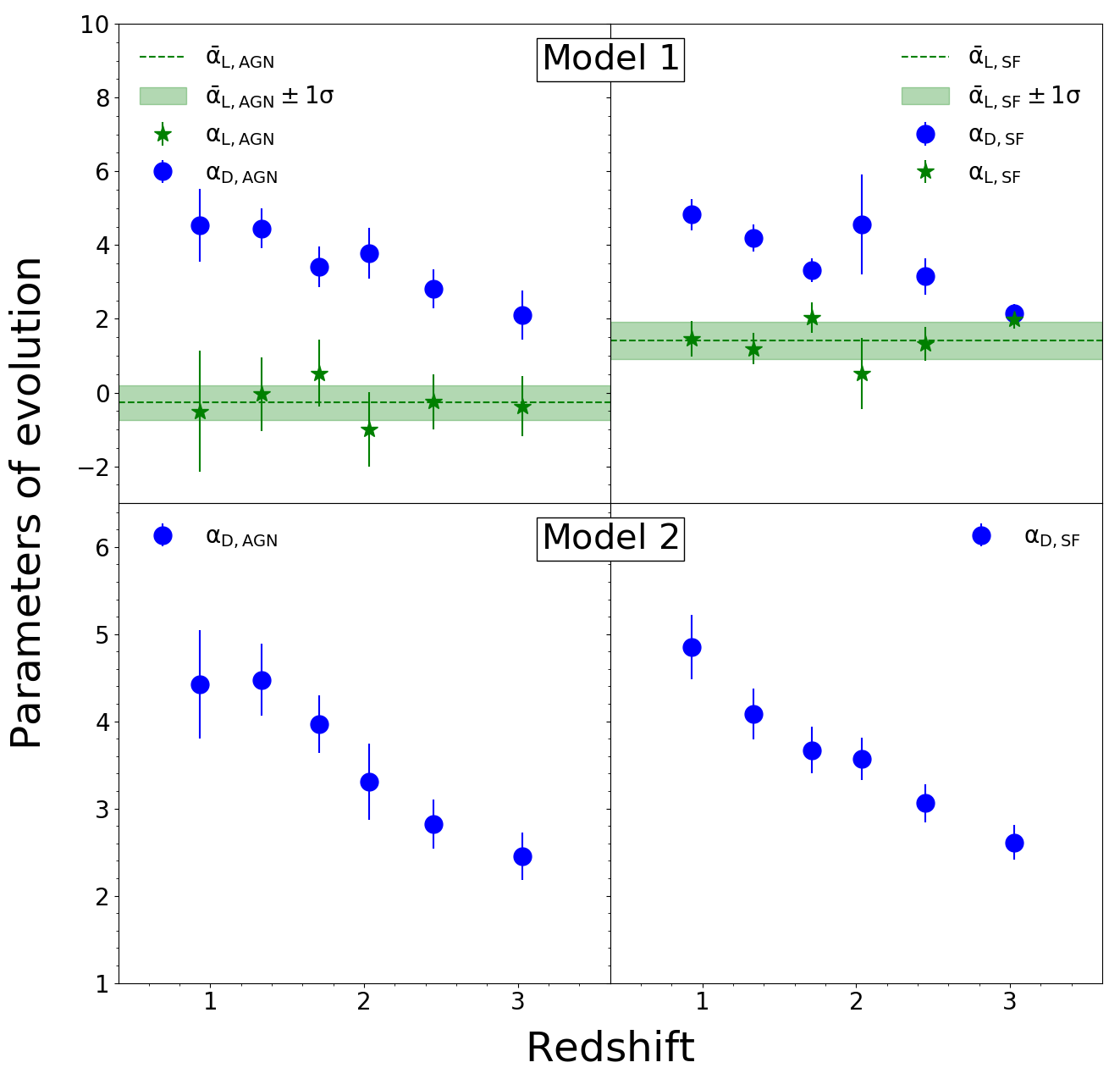}
  \caption{{Parameters of evolution obtained from fitting Model 1 (upper panels) and Model 2 (lower panels). Left and right panels show the luminosity and density evolution parameters of $\mathrm{\Phi_{AGN}}$ and $\mathrm{\Phi_{SF}}$ components of $\mathrm{\Phi_{TOT}}$.}}
  \label{fig:evol_params}\vspace{1cm}
\end{figure}

\textbf{Model 1 - Luminosity and density evolution of $\mathrm{\Phi_{TOT}}$.} 
In this model we tested the evolution of all four evolution parameters simultaneously. The results are shown in the upper panels of Fig. \ref{fig:evol_params} and listed in Table \ref{tab:evolution_parameters}. We found that both $\mathrm{\alpha_{D,SF}}$ and $\mathrm{\alpha_{D,AGN}}$ density evolution parameters show a decline with increasing redshift. The luminosity evolution parameter of the AGN component, $\mathrm{\alpha_{L,AGN}}$, has a mean value of ($\mathrm{-0.27 \pm 0.46}$) and is consistent with no redshift-dependent luminosity evolution. The luminosity evolution parameter of the SF component, $\mathrm{\alpha_{L,SF}}$, is consistent with being constant with a mean value of ($\mathrm{1.42 \pm 0.51}$).

\textbf{Model 2 - Density evolution of $\mathrm{\Phi_{TOT}}$ with fixed $\mathrm{\alpha_{L,SF}}$ and $\mathrm{\alpha_{L,AGN}}$.} 
In order to simplify the fitting procedure, in this model we set the luminosity evolution parameters to the mean values found from Model 1 ($\mathrm{\alpha_{L,SF}}=1.42$, $\mathrm{\alpha_{L,AGN}}=-0.27$). 
With this model we found that the values of the density evolution parameters for both AGN and SF components of the RLF show a global decline with redshift, as shown in the lower panels of Fig. \ref{fig:evol_params} and listed in Table \ref{tab:evolution_parameters}.

We conclude that XQSOs evolve strongly with redshift in terms of the density parameters of both SF and AGN components of the RLF. In terms of luminosity evolution, parameters of both the SF and AGN components are consistent with being constant with redshift.

We note that the normalization of the radio luminosity function of optically selected quasars by \citet{kimball11} and \citet{condon13}, shown in the first panel of Figure \ref{fig:xqso_rlf}, is more than an order of magnitude lower than that constrained by fitting our data with the analytic RLF form of Eq. \ref{eq:rlf_tot_evol}. The main cause of this difference is most likely different sample selection. While optically-selected quasar samples by \citeauthor{kimball11} and \citeauthor{condon13} contain only the unobscured quasars, sampling quasars with very high bolometric luminosities, our XQSO sample contains a combination of both obscured and unobscured quasars. By examining the hardness ratios (HRs) of XQSOs (\citealt{civano16}, \citealt{marchesi16}, \citealt{pierre16}) and applying a threshold of $\mathrm{HR = -0.2}$ to separate obscured from unobscured sources (e.g., \citealt{civano12}), we found that most of XQSOs (190/274) are obscured based on their X-ray emission and most likely would not be included in samples of optically-selected quasars.

\begin{table*}
\begin{center}
\caption{{Evolution parameters derived for the two evolution models: i) Model 1 in which both AGN and SF contributions to the $\mathrm{\Phi_{TOT}}$ can vary in luminosity and density over cosmic time and ii) Model 2 in which we fix the luminosity evolution parameters ($\mathrm{\alpha_{L, AGN}}$, $\mathrm{\alpha_{L, SF}})$ to the mean value obtained for Model 1 and test density evolution parameters of AGN and SF contributions to $\mathrm{\Phi_{TOT}}$.}}
\label{tab:evolution_parameters}
\begin{tabularx}{\linewidth} [t]{c c c c c c c c c}
Redshift
& \multicolumn{5}{c}{$\text{Model 1}$}
& \multicolumn{3}{c}{$\text{Model 2}$}\\ 
& $\mathrm{\alpha_{L, SF}}$ & $\mathrm{\alpha_{D, SF}}$ & $\mathrm{\alpha_{L, AGN}}$ & $\mathrm{\alpha_{D, AGN}}$ & $\mathrm{\chi^2}$ & $\mathrm{\alpha_{D, SF}}$ & $\mathrm{\alpha_{D, AGN}}$  & $\mathrm{\chi^2}$ \\
$0.50<z<1.13$ & $1.46 \pm 0.49$ & $4.83 \pm 0.43$ & $-0.51 \pm 1.65$ & $4.53 \pm 0.99$ & $0.13$ & $4.85 \pm 0.37$ & $4.43 \pm 0.62$ & $0.71$ \\
$1.13<z<1.48$ & $1.19 \pm 0.43$ & $4.19 \pm 0.36$ & $-0.04 \pm 1.00$ & $4.45 \pm 0.55$ & $0.33$ & $4.08 \pm 0.29$ & $4.48 \pm 0.41$ & $1.43$ \\
$1.48<z<1.85$ & $2.03 \pm 0.41$ & $3.32 \pm 0.31$ & $0.52 \pm 0.91$ & $3.42 \pm 0.55$ & $0.07$ & $3.67 \pm 0.27$ & $3.97 \pm 0.33$ & $1.60$ \\
$1.85<z<2.23$ & $0.52 \pm 0.96$ & $4.56 \pm 1.36$ & $-0.99 \pm 1.01$ & $3.77 \pm 0.68$ & $0.04$ & $3.57 \pm 0.24$ & $3.31 \pm 0.44$ & $2.08$ \\
$2.23<z<2.69$ & $1.32 \pm 0.47$ & $3.16 \pm 0.49$ & $-0.24 \pm 0.75$ & $2.82 \pm 0.53$ & $<0.01$ & $3.06 \pm 0.22$ & $2.82 \pm 0.28$ & $0.03$ \\
$2.69<z<3.75$ & $1.98 \pm 0.24$ & $2.15 \pm 0.24$ & $-0.37 \pm 0.81$ & $2.10 \pm 0.66$ & $0.50$ & $2.61 \pm 0.20$ & $2.45 \pm 0.27$ & $4.37$ \\

\end{tabularx}
\end{center}
\end{table*}

\section{Discussion} \label{sec:discussion}

In this section, we discuss the possible contributions to the detected radio emission: star-forming processes within the host galaxy in Sect. \ref{sec:host_emission} and AGN activity in Sect. \ref{sec:agn_emission}. In Sect. \ref{sec:origin_of_radio} we define thresholds in 1.4 GHz radio luminosity below and above which more than 80$\%$ of the XQSOs are dominated by the star formation and AGN-related radio emission, respectively. This enables us to disclose the true composite nature of the XQSOs in which both star formation and AGN activity make a significant contribution to the radio luminosity.

\subsection{Host galaxies of XQSOs} \label{sec:host_emission}

The sample of XQSOs studied in this work was drawn from the moderate-to-high radiative luminosity AGN sample in COSMOS and high-excitation radio galaxy sample of AGN in XXL-S. Host galaxy properties of these AGN have been studied by \citeauthor{delvecchio17} (\citeyear{delvecchio17}; COSMOS) and in Paper XXXI (XXL-S). Both studies found that the host galaxies of these AGN have star formation rates (SFRs) and stellar masses consistent with typical main-sequence galaxies. 

The same conclusion was reached by \citet{stanley15} from an analysis of a sample of $\sim$2000 X-ray selected AGN ($\mathrm{10^{42}\ erg\,s^{-1} <L_{2-8\ keV}<10^{45.5}\ erg\,s^{-1}}$) detected in the Chandra Deep Field North and South and the COSMOS field. They studied the distribution of star formation-related IR luminosity and X-ray luminosity over $\mathrm{0.2<z<2.5}$, as proxies of SFRs and AGN activity, respectively. They found a strong evolution of their average $\mathrm{L_{IR, SF}}$ with redshift, similar to that observed for the main-sequence galaxies. These results were confirmed by multiple other studies of quasar and AGN host galaxy properties (e.g., \citealt{rosario12}, \citealt{rosario13}, \citealt{lanzuisi17}, \citealt{stanley17}, \citealt{suh17}, \citealt{suh19}). {However, the relationship between the $\mathrm{L_{IR, SF}}$ and $\mathrm{L_{X}[2-8\ keV]}$ appears flat as a result of different timescales of star-forming processes and AGN activity. }

\begin{figure}[!h]
\includegraphics[width=\linewidth]{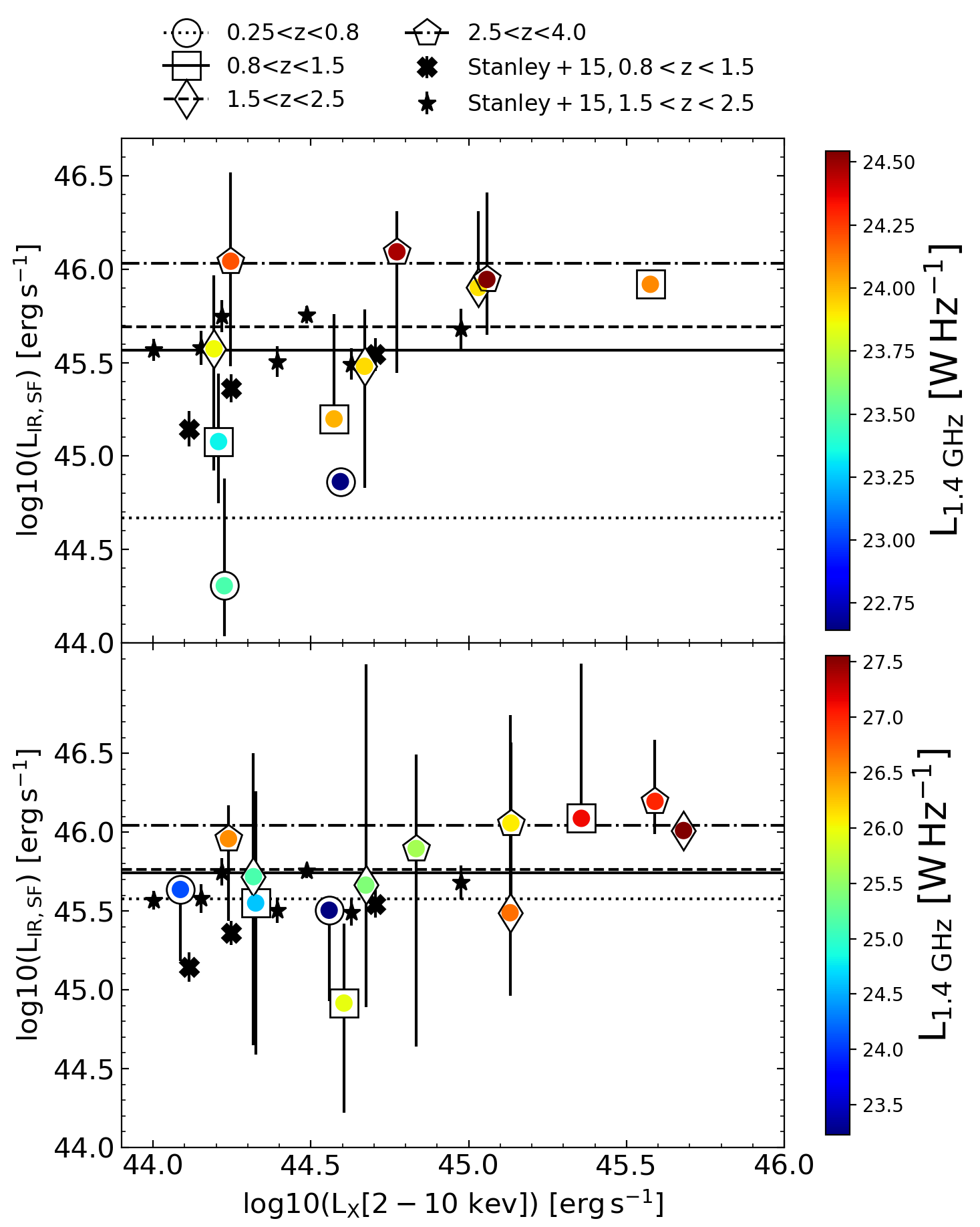}
\caption{$\mathrm{L_{IR, SF}}$ vs. $\mathrm{L_{X}[2-10\ keV]}$ of the XQSOs separated in four redshift bins, color-coded by their median 1.4 GHz radio luminosity. The upper and lower panels show XQSOs from the COSMOS and XXL-S fields, respectively. Lines show the mean value of $\mathrm{L_{IR, SF}}$ in different redshift bins, as indicated in legend. We note that the two color-coded bars correspond to very different ranges in $\mathrm{L_{1.4\ GHz}}$ in COSMOS and XXL-S.}
\label{fig:ir_x_comparison}
\end{figure}

In Fig. \ref{fig:ir_x_comparison}, we show the $\mathrm{log_{10}(L_{IR, SF})}$ (derived from the rest-frame $\mathrm{8-1000\ \mu m}$ luminosity) as a function of $\mathrm{log_{10}(L_{X}[2-10\ keV])}$ of our XQSOs for both COSMOS (upper panel) and XXL-S (lower panel) sources, binned by X-ray luminosity (0.5 dex bin width). The dashed lines show the mean $\mathrm{L_{IR, SF}}$ for all the sources per redshift bin. We compare our results with those for X-ray selected AGN by \citet{stanley15}, by converting their $\mathrm{[2-8\ keV]}$ X-ray luminosities into $\mathrm{[2-10\ keV]}$ X-ray luminosities, assuming photon index value of 1.9. In Fig. \ref{fig:ir_x_comparison} we show only the results by \citeauthor{stanley15} in matching redshift ranges ($\mathrm{0.8<z<1.5}$ and $\mathrm{1.5<z<2.5}$) and above the X-ray threshold applied to define our sample of XQSOs ($\mathrm{L_{X}[2-10\ keV] = 10^{44}\ erg\,s^{-1}}$).

For the COSMOS XQSOs, we find a rapid redshift evolution of the mean $\mathrm{L_{IR, SF}}$ (lines in the upper panel of Fig. \ref{fig:ir_x_comparison}) with an average increase of IR luminosities by a factor $\sim$3.8 from one redshift bin to the next one, in rough agreement with results by \citet{stanley15}. However, XXL-S XQSOs show a slower evolution of the $\mathrm{L_{IR, SF}}$ with an average increase by a factor of $\sim$1.5. XXL-S XQSOs have higher mean values of $\mathrm{L_{IR, SF}}$ in the first redshift bin ($\mathrm{0.25<z<0.8}$) than the COSMOS XQSOs. {The observed difference most likely arises due to different sensitivities of the radio data in the COSMOS and XXL-S field, with the COSMOS data-set being more sensitive to fainter radio sources.} XQSOs with $\mathrm{L_{X}[2-10\ keV] > 10^{45}\ erg\,s^{-1}}$ tend to have higher $\mathrm{L_{IR, SF}}$ than the lower $\mathrm{L_{X}[2-10\ keV]}$ XQSOs. There is a hint that higher redshift bins contain sources with higher median values of 1.4 GHz radio luminosities, both in COSMOS and XXL-S.

\begin{figure}[!h]
\includegraphics[width=\linewidth]{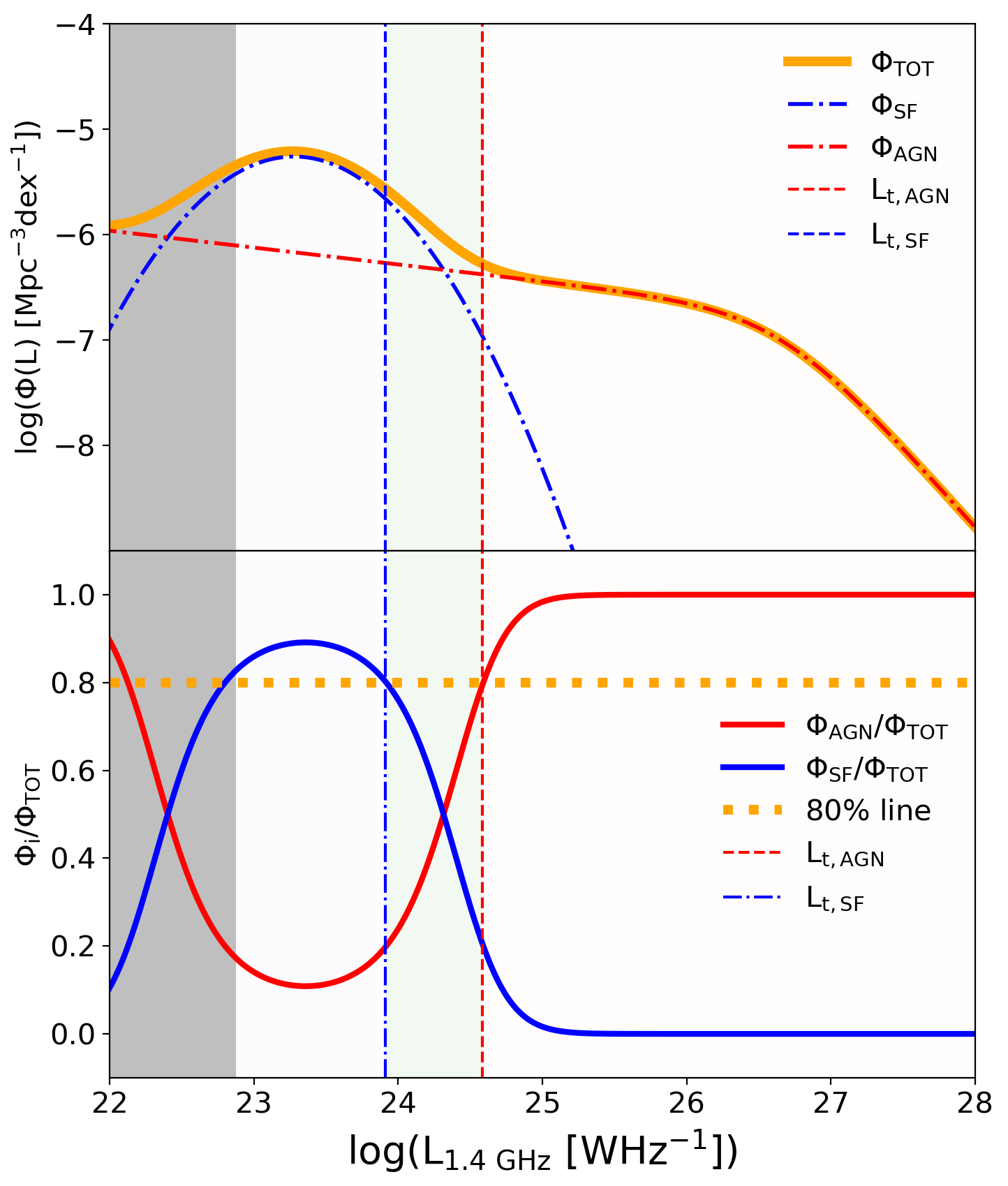}
\caption{Example of how luminosity thresholds $\mathrm{L_{t, SF}}$ and $\mathrm{L_{t, AGN}}$ (blue and red dashed lines, respectively) were defined. \textit{Top:} Radio luminosity function ($\mathrm{\Phi_{TOT}}$; orange line), constrained by the data at $\mathrm{2.23<z<2.69}$, is separated into a lower-luminosity "bump" ($\mathrm{\Phi_{SF}}$; dot-dashed blue line) and the higher-luminosity double power-law ($\mathrm{\Phi_{AGN}}$; dot-dashed red line). \textit{Bottom}: Ratios of the SF ($\mathrm{\Phi_{SF}}$) and AGN ($\mathrm{\Phi_{AGN}}$) contributions to the total RLF ($\mathrm{\Phi_{TOT}}$) as a function of $\mathrm{L_{1.4\ GHz}}$ are shown by blue and red lines, respectively. $\mathrm{L_{t, SF}}$ and $\mathrm{L_{t, AGN}}$ are values of $\mathrm{L_{1.4\ GHz}}$ below and above which more than 80$\%$ of sources used to constrain the RLF are dominated by star formation and AGN-related radio emission, respectively. In both panels, the gray area shows the region of the plot in which the curves are just an extrapolation of the analytic forms of RLFs below the detection limits of radio data used to constrain them.}
\label{fig:lt_definition}
\end{figure}

We can use the mean values of $\mathrm{L_{IR, SF}}$ to make a rough qualitative estimate of the contribution of star formation to the total radio luminosity, $\mathrm{L_{1.4\ GHz, SF}}$. The mean $\mathrm{L_{IR, SF}}$ values can be scaled to SFRs using \citet{kennicutt98} relation, which can then be used to calculate the expected radio emission via the redshift dependent $\mathrm{q_{TIR}}$ parameter by \citet{delhaize17} (assuming a radio spectral index $\mathrm{\alpha=-0.7}$). Estimated values of $\mathrm{L_{1.4\ GHz, SF}}$ for the COSMOS XQSOs increase with redshift, ranging from $\sim$$\mathrm{3\times10^{22}\ W\,Hz^{-1}}$ to $\sim$$\mathrm{2\times10^{24}\ W\,Hz^{-1}}$. Compared to the median total radio luminosity of COSMOS XQSOs per redshift bin, star formation can make a significant contribution to the total radio luminosity and even be a dominant source of radio emission. The estimates of the XXL-S XQSO $\mathrm{L_{1.4\ GHz, SF}}$ are comparable to those found for COSMOS XQSOs and show an increasing trend, ranging from  $\sim$$\mathrm{2\times10^{23}\ W\,Hz^{-1}}$ to $\sim$$\mathrm{2\times10^{24}\ W\,Hz^{-1}}$. 

\subsection{{AGN-related radio emission in XQSOs}} \label{sec:agn_emission}

The extent to which AGN-related radio emission can contribute to the total observed radio emission in AGN and quasars has been a subject of several recent studies. 
\citet{white15} studied the properties of 74 RQ quasars (RQQs) selected in optical and NIR bands within a 1 $\mathrm{deg^2}$ area covered with the VIDEO Survey. They used the stacking technique to extract the information below the detection limit of the VLA-VIRMOS Deep Field survey (\citealt{bondi03}) at 1.4 GHz, finding evidence of radio emission in their quasar sample. By comparing the radio-based SFRs (\citealt{yun01}) and those based on the assumption that quasar host galaxies lie on the main sequence of SFGs (\citealt{whitaker12}), they found a mismatch in distributions which they attributed to a significant amount of AGN-related contribution to the total radio flux in these sources. They argued that the primary origin of the radio emission in sources at $\mathrm{S_{1.4\ GHz} < 1\ mJy}$ are BH accretion processes. The same conclusion was reached by \citet{white17} who studied a sample of 70 RQQs from \textit{Spitzer-Herschel} Active Galaxy Survey at $\mathrm{0.9<z<1.1}$. They performed targeted observations of RQQs with the VLA at 1.5 GHz, detecting 35/70 of sources within their sample above $\mathrm{2 \sigma}$. For the sources with both radio and FIR detections (26/70), comparing the rest-frame $\mathrm{125\ \mu m}$ luminosity (as a tracer of the level of star formation within the host galaxy) and the 1.5 GHz radio luminosity with FIR-to-radio correlation by \citet{smith14}, they found that $\mathrm{92\%}$ of them are dominated by AGN-related radio emission. 

The above results were obtained by extracting signal below the radio detection limit and for a relatively small sample of optically very bright RQQs at $\mathrm{z\sim1}$ with targeted radio observations. As we show in the next section, by combining deep radio data within the COSMOS field with shallower radio data gathered over the larger XXL-S field we are able to study the origin of the radio emission in XQSOs via the shape of the radio luminosity function over five orders of magnitude in $\mathrm{L_{1.4\ GHz}}$. This enables us to capture the true composite nature of quasars in which both an AGN and star formation can make a noticeable contribution to the observed radio emission.

\subsection{Origin of XQSO radio emission} \label{sec:origin_of_radio}

From our RLF analysis, we found evidence of star-forming processes operating in galaxies hosting XQSOs. This means that star formation-related radio emission is expected to contribute to the total radio emission detected in our XQSOs at radio luminosities below $\mathrm{L_{1.4\ GHz} \sim 10^{24}\ W\,Hz^{-1}}$.

To define redshift-dependent thresholds in $\mathrm{L_{1.4\ GHz}}$ below and above which the dominant contribution to the RLF arises from the star formation and AGN activity, we calculated ratios of $\mathrm{\Phi_{SF}}$-to-$\mathrm{\Phi_{TOT}}$ and $\mathrm{\Phi_{AGN}}$-to-$\mathrm{\Phi_{TOT}}$ as a function of 1.4 GHz luminosity, as shown in Fig. \ref{fig:lt_definition}. Luminosity dependent values of $\mathrm{\Phi_{SF}}$, $\mathrm{\Phi_{AGN}}$ and $\mathrm{\Phi_{TOT}}$ (see Sect. \ref{sec:rlf_evol}) were calculated at the median redshift in six redshift bins using the evolution parameters calculated for the luminosity and density evolution model of $\mathrm{\Phi_{TOT}}$ (see Model 1 in Table \ref{tab:evolution_parameters}). For each redshift bin we defined two threshold luminosities, $\mathrm{L_{t, SF}}$ and $\mathrm{L_{t, AGN}}$, where $\mathrm{L_{t, SF}}$ is the luminosity below which we expect $\geq$80$\%$ of sources contributing to the RLF to be dominated by star formation-related radio emission and $\mathrm{L_{t, AGN}}$ is the luminosity above which $\geq$80$\%$ of sources are expected to have radio emission dominated by AGN activity. 
Sources with luminosities $\mathrm{L_{t,SF}\leq L_{1.4\ GHz} \leq L_{t,AGN}}$ are expected to be those in which both star formation and AGN activity have a significant contribution to the total radio emission. The numbers of COSMOS and XXL-S sources within each radio luminosity range are listed in Table \ref{tab:numbers}. 

\begin{figure}[!h]
\includegraphics[width=\linewidth]{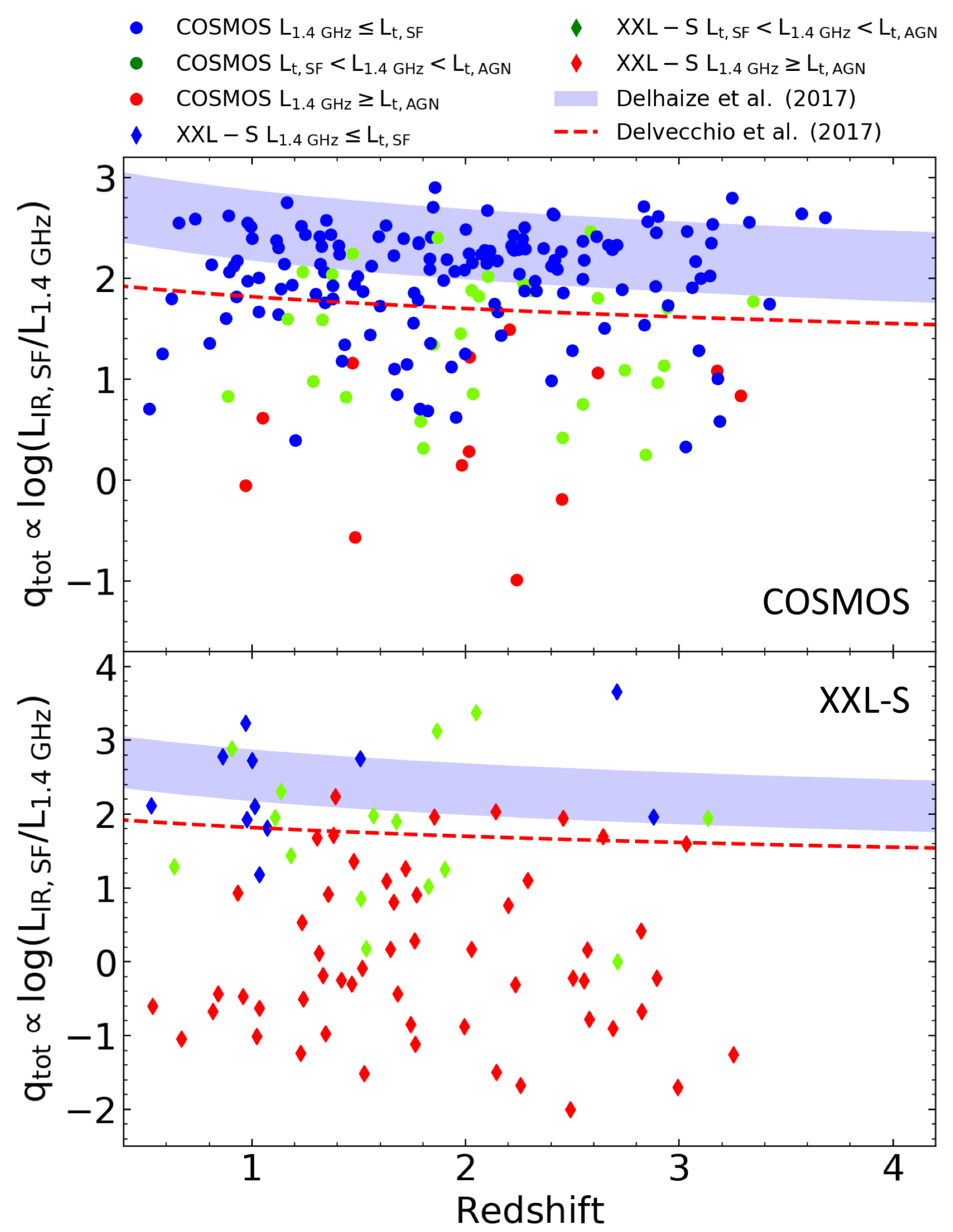}
\caption{$\mathrm{q_{TOT}}$ vs. redshift of COSMOS (upper panel) and XXL-S (lower panel) XQSOs. Sources with $\mathrm{L_{1.4\ GHz}\leq L_{t, SF}}$, $\mathrm{L_{t, SF} < L_{1.4\ GHz} < L_{t, AGN}}$ and $\mathrm{L_{1.4\ GHz}\geq L_{t, AGN}}$ are shown with blue, green and red symbols respectively. {The red dashed line shows the radio-excess threshold defined by \citet{delvecchio17}. The blue shaded area shows the "SFG locus" by \citet{delhaize17}.}}
\label{fig:z_vs_q}
\end{figure}

In Fig. \ref{fig:z_vs_q}, we show XQSOs in $\mathrm{q_{TOT}}$ vs. z plane. The infrared-radio correlation found for the sample of SFGs, expressed via the $\mathrm{q_{TIR}}$ parameter, is a valuable tool for quantifying the amount of radio emission arising from star formation within the host galaxy (\citealt{delhaize17}, \citealt{ceraj18}). If the radio luminosity of the source is dominated by the star formation-related emission, it would populate the so-called "SFG locus" of the $\mathrm{q_{TOT}}$ vs. $\mathrm{z}$ plane. In Fig. \ref{fig:z_vs_q} we show this locus using the redshift-evolving $\mathrm{q_{TIR}\pm0.35}$ by \citet{delhaize17}. Below this locus, the radio excess due to the AGN-related radio emission increases with decreasing $\mathrm{q_{TOT}}$. We found that most COSMOS XQSOs are located within or near the "SFG locus", while the XXL-S XQSOs mostly show significant radio excess. This indicates that a large fraction of COSMOS XQSOs have their emission dominated by star formation, while most of the XXL-S XQSOs radio emission comes from AGN activity.

We further separated COSMOS and XXL-S XQSOs with respect to the radio luminosity thresholds $\mathrm{L_{t, SF}}$ and $\mathrm{L_{t, AGN}}$ in the $\mathrm{q_{TOT}}$ vs. z plane. The number of XQSOs which are below the radio excess threshold by \citealt{delvecchio17}, $\mathrm{q_{REX}= 24-21.984\times(1+z)^{0.013}}$, are shown in parentheses in Table \ref{tab:numbers}. The results show that the majority of sources contribution to the RLF "bump" have their radio luminosity dominated by star formation-related emission, while the XQSOs at the higher-luminosity end of RLF are dominated by the AGN radio emission.

\begin{table}
\centering
\caption{Number of COSMOS and XXL-S XQSOs divided in ranges of $\mathrm{L_{1.4\ GHz}}$ defined on the basis of the thresholds $\mathrm{L_{t,SF}}$ and $\mathrm{L_{t,AGN}}$. Numbers in parentheses show the number of sources within each $\mathrm{L_{1.4\ GHz}}$ interval that would be defined as radio-excess with respect to criterion defined by \citet{delvecchio17}.}
\quad
\renewcommand{\arraystretch}{1.5}
\begin{tabular}[t]{c c c}
\hline
 & COSMOS & XXL-S\\
\hline
$\mathrm{L_{1.4\ GHz}\leq L_{t, SF}}$ & 143 (34) & 11 (1) \\
$\mathrm{L_{t, SF} < L_{1.4\ GHz} < L_{t, AGN}}$ & 30 (16) & 15 (7)\\
$\mathrm{L_{1.4\ GHz}\geq L_{t, AGN}}$ & 13 (13) & 55 (50) \\

\hline
\end{tabular}
\label{tab:numbers}
\end{table}

\begin{figure}[!htbp]
  \includegraphics[width=\linewidth]{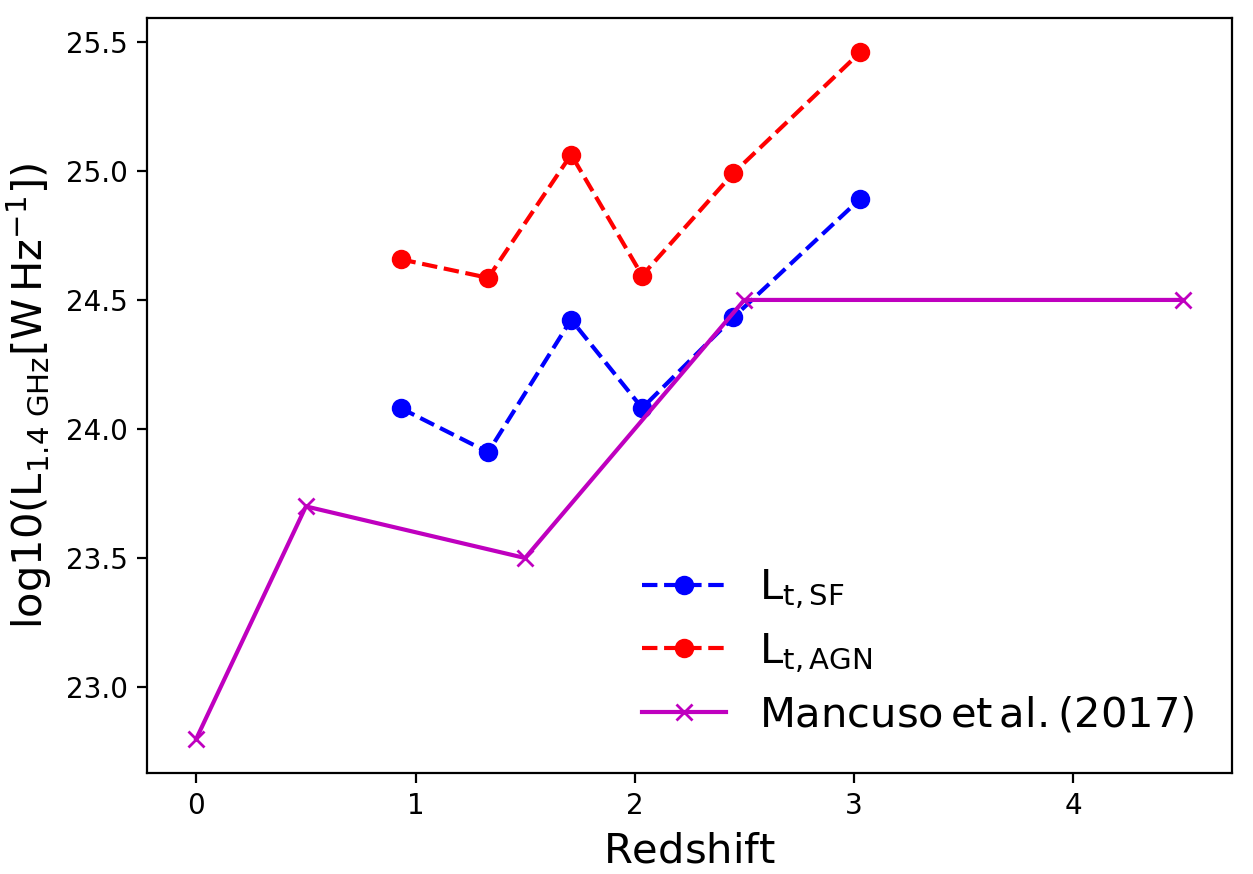}
  \caption{1.4 GHz radio luminosity thresholds vs. redshift. Blue and red dots and dotted lines show the redshift evolution of thresholds below and above more than $\mathrm{80\,\%}$ of sources contributing to the XQSO RLF are dominated by star formation and AGN activity, respectively. Magenta symbols and line are taken from \citet{mancuso17} and represent the threshold at which SFGs and RQ AGN change dominance in their RLFs.}
  \label{fig:lth_comp}\vspace{1cm}
\end{figure}

Threshold luminosities, $\mathrm{L_{t, SF}}$ and $\mathrm{L_{t, AGN}}$, shown in Fig. \ref{fig:xqso_rlf} and listed in Table \ref{tab:lt}, appear to evolve with redshift from lower to higher 1.4 GHz luminosities, as shown in Fig. \ref{fig:lth_comp}. A similar trend has been found in models of galaxy evolution by \citet{mancuso17}, who studied evolutionary tracks of SFGs and radio-silent (RS), RQ and RL AGN. They found that the 1.4 GHz luminosity threshold, at which the dominance between SFGs and RS+RQ AGN changes in RLFs, increases with redshift. They argue that this evolution is result of strong cosmic evolution of the star formation in host galaxies. A similar conclusion can be drawn for the evolution of $\mathrm{L_{t, SF}}$ and $\mathrm{L_{t, AGN}}$ in RLFs of XQSOs. As demonstrated in Sect. \ref{sec:host_emission}, star formation-related IR luminosity evolves with redshift, which indicates that strong evolution of radio emission due to star-forming processes should be expected, as is confirmed in Sect. \ref{sec:rlf_evol}.

While the majority of XQSOs at $\mathrm{L_{t, SF}>L_{1.4\ GHz}}$ are dominated by the radio emission of star forming origin, the AGN-related radio emission might still be present at lower levels. As discussed by \citealt{lb08}, radio-quiet quasars selected from the Palomar-Green (PG) Bright Quasar Survey ($\mathrm{z\leq0.5}$, \citealt{sg83}) follow a tight correlation between their radio and X-ray luminosities. This correlation seems to be extension of the so-called G{\"u}del-Benz correlation found to exist between the quiescent radio and X-ray emission of coronally active stars (\citealt{gb93}) in which both these emissions have coronal origin. By comparing the radio and X-ray luminosities of XQSOs to the \citeauthor{lb08} relation, we found that at least some of the $\mathrm{L_{t, SF}>L_{1.4\ GHz}}$ XQSOs could have the radio emission of coronal origin.

\begin{table}
\centering
\caption{Luminosity thresholds $\mathrm{L_{t, SF}}$ and $\mathrm{L_{t, AGN}}$ constrained using XQSO RLFs.}
\quad
\renewcommand{\arraystretch}{1.5}
\begin{tabular}[t]{c c c}
\hline
$\mathrm{z_{med}}$ & $\mathrm{log_{10} (L_{t, SF})}$ & $\mathrm{log_{10}(L_{t, AGN})}$\\
\hline
0.93 & 24.08 & 24.66 \\
1.33 & 23.91 & 24.59 \\
1.71 & 24.42 & 25.06 \\
2.03 & 24.09 & 24.60 \\
2.45 & 24.43 & 24.99 \\
3.03 & 24.89 & 25.46 \\

\hline
\end{tabular}
\label{tab:lt}
\end{table}

\section{Summary and conclusions} \label{sec:sac}

We studied the origin of the radio emission of a sample of 274 radio and X-ray selected quasars (XQSOs) at $\mathrm{0.25<z<6.1}$ detected at 3 GHz and 2.1 GHz in the COSMOS and XXL-S fields, respectively. Using the multiwavelength data available for these sources, we examined four different criteria of radio loudness and found that $\mathrm{18-73 \%}$ of XQSOs are selected as radio-loud. This disagreement arises both from different flux density thresholds of COSMOS and XXL-S radio surveys and from the fact that the four radio loudness criteria capture different physical processes related to the AGN activity (radio luminosity, B-band luminosity, hard X-ray luminosity) and star-forming processes within the AGN host galaxies (star formation-related infrared luminosity). These different fractions of XQSOs selected as radio-loud and radio-quiet indicate that another approach is needed to disclose the true origin of radio emission in XQSOs.

We constructed 1.4 GHz radio luminosity functions (RLFs) for a subsample of 267 XQSOs in six redshift bins between $\mathrm{0.5<z<3.75}$. The lower-luminosity end of the RLF at all redshifts has a higher normalization than expected just from AGN-related radio emission and manifests as a "bump" in the RLF. To constrain the shape and the evolution of the XQSOs, we used the analytic representation of the local radio luminosity function from the literature. The RLF employed is a combination of parabolic and double power-law functions constraining the lower and higher-luminosity end of RLF, respectively. We found that the lower- and higher-luminosity ends evolves significantly with redshift in terms of density. The luminosity evolution parameters of both the lower- and higher-luminosity end are consistent with being constant with redshift.

We used the RLF in different redshift bins to define thresholds in 1.4 GHz radio luminosity below and above which more than $\mathrm{80 \%}$ of sources contributing to the RLF are expected to be dominated by star formation and AGN activity, respectively. To test this, we exploited the so-called $\mathrm{q_{TOT}}$ parameter, which is proportional to the logarithm of the ratio of IR luminosity due to star formation and 1.4 GHz luminosity. We confirmed that the majority of sources contributing to the "bump" of the RLF are dominated by star formation-related radio emission, while the sources contributing to higher-luminosity end of RLF are dominated by AGN-related radio emission. Our result exposes the dichotomy of the processes behind the XQSO radio emission: star formation dominating the radio emission of the lower radio luminosity and AGN activity dominating the higher radio luminosity.

\begin{acknowledgements}
The authors are grateful to the anonymous referee for helpful and insightful comments which helped to improve the content of this article. Based on observations obtained with XMM-Newton, an ESA science mission with instruments and contributions directly funded by ESA Member States and NASA. LC, VS and KT acknowledge support from the European Union's Seventh Framework program under grant agreement 337595 (ERC Starting Grant, 'CoSMass'). ID is supported by the European Union's Horizon 2020 research and innovation program under the Marie Sk\l{}odowska-Curie grant agreement No 788679. MN acknowledges support from the ERC Advanced Grant 740246 (Cosmic Gas). The Saclay group acknowledges long-term support from the Centre National d'Etudes Spatiales (CNES).
XXL is an international project based around an XMM Very Large Programme surveying two $\mathrm{25\,deg^2}$ extragalactic fields at a depth of $\sim$6$\mathrm{\times 10-15\,erg\,cm^{-2}\,s^{-1}}$ in the [0.5-2] keV band for point-like sources. The XXL website is http://irfu.cea.fr/xxl. Multi-band information and spectroscopic follow-up of the X-ray sources are obtained through a number of survey programmes, summarised at http://xxlmultiwave.pbworks.com/. 
\end{acknowledgements}

\bibliographystyle{aa} 
\bibliography{bibliography}

\begin{thebibliography}{64}
\expandafter\ifx\csname natexlab\endcsname\relax\def\natexlab#1{#1}\fi

\bibitem[{{Balokovi{\'c}} {et~al.}(2012){Balokovi{\'c}}, {Smol{\v c}i{\'c}},
  {Ivezi{\'c}}, {Zamorani}, {Schinnerer}, \& {Kelly}}]{balokovic12}
{Balokovi{\'c}}, M., {Smol{\v c}i{\'c}}, V., {Ivezi{\'c}}, {\v Z}., {et~al.}
  2012, \apj, 759, 30

\bibitem[{{Bondi} {et~al.}(2003){Bondi}, {Ciliegi}, {Zamorani}, {Gregorini},
  {Vettolani}, {Parma}, {de Ruiter}, {Le Fevre}, {Arnaboldi}, {Guzzo},
  {Maccagni}, {Scaramella}, {Adami}, {Bardelli}, {Bolzonella}, {Bottini},
  {Cappi}, {Foucaud}, {Franzetti}, {Garilli}, {Gwyn}, {Ilbert}, {Iovino}, {Le
  Brun}, {Marano}, {Marinoni}, {McCracken}, {Meneux}, {Pollo}, {Pozzetti},
  {Radovich}, {Ripepi}, {Rizzo}, {Scodeggio}, {Tresse}, {Zanichelli}, \&
  {Zucca}}]{bondi03}
{Bondi}, M., {Ciliegi}, P., {Zamorani}, G., {et~al.} 2003, \aap, 403, 857

\bibitem[{{Butler} {et~al.}(2018{\natexlab{a}}){Butler}, {Huynh}, {Delhaize},
  {Smol{\v c}i{\'c}}, {Kapi{\'n}ska}, {Milakovi{\'c}}, {Novak}, {Baran},
  {O'Brien}, {Chiappetti}, {Desai}, {Fotopoulou}, {Horellou}, {Lidman}, \&
  {Pierre}}]{butler18a}
{Butler}, A., {Huynh}, M., {Delhaize}, J., {et~al.} 2018{\natexlab{a}}, \aap,
  620, A3, (XXL Paper XVIII)

\bibitem[{{Butler} {et~al.}(2018{\natexlab{b}}){Butler}, {Huynh}, {Delvecchio},
  {Kapi{\'n}ska}, {Ciliegi}, {Jurlin}, {Delhaize}, {Smol{\v c}i{\'c}}, {Desai},
  {Fotopoulou}, {Lidman}, {Pierre}, \& {Plionis}}]{butler18b}
{Butler}, A., {Huynh}, M., {Delvecchio}, I., {et~al.} 2018{\natexlab{b}}, \aap,
  620, A16, (XXL Paper XXXI)

\bibitem[{{Butler}(2019)}]{butler_phdt}
{Butler}, A.~R. 2019, PhD thesis, University of Western Australia

\bibitem[{{Ceraj} {et~al.}(2018){Ceraj}, {Smol{\v c}i{\'c}}, {Delvecchio},
  {Novak}, {Zamorani}, {Delhaize}, {Schinnerer}, {Vardoulaki}, \& {Herrera
  Ruiz}}]{ceraj18}
{Ceraj}, L., {Smol{\v c}i{\'c}}, V., {Delvecchio}, I., {et~al.} 2018, \aap,
  620, A192

\bibitem[{{Ciliegi} {et~al.}(2018){Ciliegi}, {Jurlin}, {Butler}, {Delhaize},
  {Fotopoulou}, {Huynh}, {Iovino}, {Smol{\v c}i{\'c}}, {Chiappetti}, \&
  {Pierre}}]{ciliegi18}
{Ciliegi}, P., {Jurlin}, N., {Butler}, A., {et~al.} 2018, \aap, 620, A11, (XXL
  Paper XXVI)

\bibitem[{{Cirasuolo} {et~al.}(2003){Cirasuolo}, {Celotti}, {Magliocchetti}, \&
  {Danese}}]{cirasuolo03}
{Cirasuolo}, M., {Celotti}, A., {Magliocchetti}, M., \& {Danese}, L. 2003,
  \mnras, 346, 447

\bibitem[{{Civano} {et~al.}(2012){Civano}, {Elvis}, {Brusa}, {Comastri},
  {Salvato}, {Zamorani}, {Aldcroft}, {Bongiorno}, {Capak}, {Cappelluti},
  {Cisternas}, {Fiore}, {Fruscione}, {Hao}, {Kartaltepe}, {Koekemoer}, {Gilli},
  {Impey}, {Lanzuisi}, {Lusso}, {Mainieri}, {Miyaji}, {Lilly}, {Masters},
  {Puccetti}, {Schawinski}, {Scoville}, {Silverman}, {Trump}, {Urry},
  {Vignali}, \& {Wright}}]{civano12}
{Civano}, F., {Elvis}, M., {Brusa}, M., {et~al.} 2012, \apjs, 201, 30

\bibitem[{{Civano} {et~al.}(2016){Civano}, {Marchesi}, {Comastri}, {Urry},
  {Elvis}, {Cappelluti}, {Puccetti}, {Brusa}, {Zamorani}, {Hasinger},
  {Aldcroft}, {Alexander}, {Allevato}, {Brunner}, {Capak}, {Finoguenov},
  {Fiore}, {Fruscione}, {Gilli}, {Glotfelty}, {Griffiths}, {Hao}, {Harrison},
  {Jahnke}, {Kartaltepe}, {Karim}, {LaMassa}, {Lanzuisi}, {Miyaji}, {Ranalli},
  {Salvato}, {Sargent}, {Scoville}, {Schawinski}, {Schinnerer}, {Silverman},
  {Smolcic}, {Stern}, {Toft}, {Trakhtenbrot}, {Treister}, \&
  {Vignali}}]{civano16}
{Civano}, F., {Marchesi}, S., {Comastri}, A., {et~al.} 2016, \apj, 819, 62

\bibitem[{{Condon}(1992)}]{condon92}
{Condon}, J.~J. 1992, \araa, 30, 575

\bibitem[{{Condon} {et~al.}(2002){Condon}, {Cotton}, \& {Broderick}}]{condon02}
{Condon}, J.~J., {Cotton}, W.~D., \& {Broderick}, J.~J. 2002, \aj, 124, 675

\bibitem[{{Condon} {et~al.}(1998){Condon}, {Cotton}, {Greisen}, {Yin},
  {Perley}, {Taylor}, \& {Broderick}}]{condon98}
{Condon}, J.~J., {Cotton}, W.~D., {Greisen}, E.~W., {et~al.} 1998, \aj, 115,
  1693

\bibitem[{{Condon} {et~al.}(2013){Condon}, {Kellermann}, {Kimball},
  {Ivezi{\'c}}, \& {Perley}}]{condon13}
{Condon}, J.~J., {Kellermann}, K.~I., {Kimball}, A.~E., {Ivezi{\'c}}, {\v Z}.,
  \& {Perley}, R.~A. 2013, \apj, 768, 37

\bibitem[{{Condon} {et~al.}(1980){Condon}, {Odell}, {Puschell}, \&
  {Stein}}]{condon80}
{Condon}, J.~J., {Odell}, S.~L., {Puschell}, J.~J., \& {Stein}, W.~A. 1980,
  \nat, 283, 357

\bibitem[{{Condon} {et~al.}(1981){Condon}, {Odell}, {Puschell}, \&
  {Stein}}]{condon81}
{Condon}, J.~J., {Odell}, S.~L., {Puschell}, J.~J., \& {Stein}, W.~A. 1981,
  \apj, 246, 624

\bibitem[{{Delhaize} {et~al.}(2017){Delhaize}, {Smol{\v c}i{\'c}},
  {Delvecchio}, {Novak}, {Sargent}, {Baran}, {Magnelli}, {Zamorani},
  {Schinnerer}, {Murphy}, {Aravena}, {Berta}, {Bondi}, {Capak}, {Carilli},
  {Ciliegi}, {Civano}, {Ilbert}, {Karim}, {Laigle}, {Le F{\`e}vre}, {Marchesi},
  {McCracken}, {Salvato}, {Seymour}, \& {Tasca}}]{delhaize17}
{Delhaize}, J., {Smol{\v c}i{\'c}}, V., {Delvecchio}, I., {et~al.} 2017, \aap,
  602, A4

\bibitem[{{Delvecchio} {et~al.}(2017){Delvecchio}, {Smol{\v c}i{\'c}},
  {Zamorani}, {Lagos}, {Berta}, {Delhaize}, {Baran}, {Alexander}, {Rosario},
  {Gonzalez-Perez}, {Ilbert}, {Lacey}, {Le F{\`e}vre}, {Miettinen}, {Aravena},
  {Bondi}, {Carilli}, {Ciliegi}, {Mooley}, {Novak}, {Schinnerer}, {Capak},
  {Civano}, {Fanidakis}, {Herrera Ruiz}, {Karim}, {Laigle}, {Marchesi},
  {McCracken}, {Middleberg}, {Salvato}, \& {Tasca}}]{delvecchio17}
{Delvecchio}, I., {Smol{\v c}i{\'c}}, V., {Zamorani}, G., {et~al.} 2017, \aap,
  602, A3

\bibitem[{{Gehrels}(1986)}]{gehrels86}
{Gehrels}, N. 1986, \apj, 303, 336

\bibitem[{{Goldschmidt} {et~al.}(1999){Goldschmidt}, {Kukula}, {Miller}, \&
  {Dunlop}}]{gold99}
{Goldschmidt}, P., {Kukula}, M.~J., {Miller}, L., \& {Dunlop}, J.~S. 1999,
  \apj, 511, 612

\bibitem[{{Guedel} \& {Benz}(1993)}]{gb93}
{Guedel}, M. \& {Benz}, A.~O. 1993, \apjl, 405, L63

\bibitem[{{Hao} {et~al.}(2014){Hao}, {Sargent}, {Elvis}, {Schinnerer},
  {Zamorani}, {Ho}, {Donley}, {Civano}, {Smolcic}, {Celotti}, {Kuraszkiewicz},
  {Salvato}, {Brusa}, {Capak}, {Carilli}, {Comastri}, {Impey}, {Jahnke},
  {Koekemoer}, {Schawinski}, {Trump}, {Urry}, {Vignali}, \& {Yun}}]{hao14}
{Hao}, H., {Sargent}, M.~T., {Elvis}, M., {et~al.} 2014, arXiv e-prints
  [\eprint[arXiv]{1408.1090}]

\bibitem[{{Ivezi{\'c}} {et~al.}(2002){Ivezi{\'c}}, {Menou}, {Knapp}, {Strauss},
  {Lupton}, {Vanden Berk}, {Richards}, {Tremonti}, {Weinstein}, {Anderson},
  {Bahcall}, {Becker}, {Bernardi}, {Blanton}, {Eisenstein}, {Fan},
  {Finkbeiner}, {Finlator}, {Frieman}, {Gunn}, {Hall}, {Kim}, {Kinkhabwala},
  {Narayanan}, {Rockosi}, {Schlegel}, {Schneider}, {Strateva}, {SubbaRao},
  {Thakar}, {Voges}, {White}, {Yanny}, {Brinkmann}, {Doi}, {Fukugita},
  {Hennessy}, {Munn}, {Nichol}, \& {York}}]{ivezic02}
{Ivezi{\'c}}, {\v Z}., {Menou}, K., {Knapp}, G.~R., {et~al.} 2002, \aj, 124,
  2364

\bibitem[{{Kellermann} {et~al.}(1989){Kellermann}, {Sramek}, {Schmidt},
  {Shaffer}, \& {Green}}]{kellermann89}
{Kellermann}, K.~I., {Sramek}, R., {Schmidt}, M., {Shaffer}, D.~B., \& {Green},
  R. 1989, \aj, 98, 1195

\bibitem[{{Kennicutt}(1998)}]{kennicutt98}
{Kennicutt}, Jr., R.~C. 1998, \apj, 498, 541

\bibitem[{{Kimball} {et~al.}(2011){Kimball}, {Kellermann}, {Condon},
  {Ivezi{\'c}}, \& {Perley}}]{kimball11}
{Kimball}, A.~E., {Kellermann}, K.~I., {Condon}, J.~J., {Ivezi{\'c}}, {\v Z}.,
  \& {Perley}, R.~A. 2011, \apjl, 739, L29

\bibitem[{{Laigle} {et~al.}(2016){Laigle}, {McCracken}, {Ilbert}, {Hsieh},
  {Davidzon}, {Capak}, {Hasinger}, {Silverman}, {Pichon}, {Coupon}, {Aussel},
  {Le Borgne}, {Caputi}, {Cassata}, {Chang}, {Civano}, {Dunlop}, {Fynbo},
  {Kartaltepe}, {Koekemoer}, {Le F{\`e}vre}, {Le Floc'h}, {Leauthaud}, {Lilly},
  {Lin}, {Marchesi}, {Milvang-Jensen}, {Salvato}, {Sanders}, {Scoville},
  {Smolcic}, {Stockmann}, {Taniguchi}, {Tasca}, {Toft}, {Vaccari}, \&
  {Zabl}}]{laigle16}
{Laigle}, C., {McCracken}, H.~J., {Ilbert}, O., {et~al.} 2016, \apjs, 224, 24

\bibitem[{{Lanzuisi} {et~al.}(2017){Lanzuisi}, {Delvecchio}, {Berta}, {Brusa},
  {Comastri}, {Gilli}, {Gruppioni}, {Marchesi}, {Perna}, {Pozzi}, {Salvato},
  {Symeonidis}, {Vignali}, {Vito}, {Volonteri}, \& {Zamorani}}]{lanzuisi17}
{Lanzuisi}, G., {Delvecchio}, I., {Berta}, S., {et~al.} 2017, \aap, 602, A123

\bibitem[{{Laor} \& {Behar}(2008)}]{lb08}
{Laor}, A. \& {Behar}, E. 2008, \mnras, 390, 847

\bibitem[{{Mancuso} {et~al.}(2017){Mancuso}, {Lapi}, {Prandoni}, {Obi},
  {Gonzalez-Nuevo}, {Perrotta}, {Bressan}, {Celotti}, \& {Danese}}]{mancuso17}
{Mancuso}, C., {Lapi}, A., {Prandoni}, I., {et~al.} 2017, \apj, 842, 95

\bibitem[{{Marchesi} {et~al.}(2016){Marchesi}, {Civano}, {Elvis}, {Salvato},
  {Brusa}, {Comastri}, {Gilli}, {Hasinger}, {Lanzuisi}, {Miyaji}, {Treister},
  {Urry}, {Vignali}, {Zamorani}, {Allevato}, {Cappelluti}, {Cardamone},
  {Finoguenov}, {Griffiths}, {Karim}, {Laigle}, {LaMassa}, {Jahnke}, {Ranalli},
  {Schawinski}, {Schinnerer}, {Silverman}, {Smolcic}, {Suh}, \&
  {Trakhtenbrot}}]{marchesi16}
{Marchesi}, S., {Civano}, F., {Elvis}, M., {et~al.} 2016, \apj, 817, 34

\bibitem[{{Marshall}(1985)}]{marshall85}
{Marshall}, H.~L. 1985, \apj, 299, 109

\bibitem[{{Mauch} \& {Sadler}(2007)}]{ms07}
{Mauch}, T. \& {Sadler}, E.~M. 2007, \mnras, 375, 931

\bibitem[{{Miller} {et~al.}(1990){Miller}, {Peacock}, \& {Mead}}]{miller90}
{Miller}, L., {Peacock}, J.~A., \& {Mead}, A.~R.~G. 1990, \mnras, 244, 207

\bibitem[{{Novak} {et~al.}(2017){Novak}, {Smol{\v c}i{\'c}}, {Delhaize},
  {Delvecchio}, {Zamorani}, {Baran}, {Bondi}, {Capak}, {Carilli}, {Ciliegi},
  {Civano}, {Ilbert}, {Karim}, {Laigle}, {Le F{\`e}vre}, {Marchesi},
  {McCracken}, {Miettinen}, {Salvato}, {Sargent}, {Schinnerer}, \&
  {Tasca}}]{novak17}
{Novak}, M., {Smol{\v c}i{\'c}}, V., {Delhaize}, J., {et~al.} 2017, \aap, 602,
  A5

\bibitem[{{Novak} {et~al.}(2018){Novak}, {Smol{\v c}i{\'c}}, {Schinnerer},
  {Zamorani}, {Delvecchio}, {Bondi}, \& {Delhaize}}]{novak18}
{Novak}, M., {Smol{\v c}i{\'c}}, V., {Schinnerer}, E., {et~al.} 2018, \aap,
  614, A47

\bibitem[{{Padovani} {et~al.}(2017){Padovani}, {Alexander}, {Assef}, {De
  Marco}, {Giommi}, {Hickox}, {Richards}, {Smol{\v c}i{\'c}}, {Hatziminaoglou},
  {Mainieri}, \& {Salvato}}]{padovani17}
{Padovani}, P., {Alexander}, D.~M., {Assef}, R.~J., {et~al.} 2017, \aapr, 25, 2

\bibitem[{{Perley} {et~al.}(2011){Perley}, {Chandler}, {Butler}, \&
  {Wrobel}}]{perley11}
{Perley}, R.~A., {Chandler}, C.~J., {Butler}, B.~J., \& {Wrobel}, J.~M. 2011,
  \apjl, 739, L1

\bibitem[{{Pierce} {et~al.}(2011){Pierce}, {Ballantyne}, \&
  {Ivison}}]{pierce11}
{Pierce}, C.~M., {Ballantyne}, D.~R., \& {Ivison}, R.~J. 2011, \apj, 742, 45

\bibitem[{{Pierre} {et~al.}(2016){Pierre}, {Pacaud}, {Adami}, {Alis},
  {Altieri}, {Baran}, {Benoist}, {Birkinshaw}, {Bongiorno}, {Bremer}, {Brusa},
  {Butler}, {Ciliegi}, {Chiappetti}, {Clerc}, {Corasaniti}, {Coupon}, {De
  Breuck}, {Democles}, {Desai}, {Delhaize}, {Devriendt}, {Dubois}, {Eckert},
  {Elyiv}, {Ettori}, {Evrard}, {Faccioli}, {Farahi}, {Ferrari}, {Finet},
  {Fotopoulou}, {Fourmanoit}, {Gandhi}, {Gastaldello}, {Gastaud},
  {Georgantopoulos}, {Giles}, {Guennou}, {Guglielmo}, {Horellou}, {Husband},
  {Huynh}, {Iovino}, {Kilbinger}, {Koulouridis}, {Lavoie}, {Le Brun}, {Le
  Fevre}, {Lidman}, {Lieu}, {Lin}, {Mantz}, {Maughan}, {Maurogordato},
  {McCarthy}, {McGee}, {Melin}, {Melnyk}, {Menanteau}, {Novak}, {Paltani},
  {Plionis}, {Poggianti}, {Pomarede}, {Pompei}, {Ponman}, {Ramos-Ceja},
  {Ranalli}, {Rapetti}, {Raychaudury}, {Reiprich}, {R{\"o}ttgering}, {Rozo},
  {Rykoff}, {Sadibekova}, {Santos}, {Sauvageot}, {Schimd}, {Sereno}, {Smith},
  {Smol{\v c}i{\'c}}, {Snowden}, {Spergel}, {Stanford}, {Surdej}, {Valageas},
  {Valotti}, {Valtchanov}, {Vignali}, {Willis}, \& {Ziparo}}]{pierre16}
{Pierre}, M., {Pacaud}, F., {Adami}, C., {et~al.} 2016, \aap, 592, A1, (XXL
  Paper I)

\bibitem[{{Pracy} {et~al.}(2016){Pracy}, {Ching}, {Sadler}, {Croom}, {Baldry},
  {Bland-Hawthorn}, {Brough}, {Brown}, {Couch}, {Davis}, {Drinkwater},
  {Hopkins}, {Jarvis}, {Jelliffe}, {Jurek}, {Loveday}, {Pimbblet}, {Prescott},
  {Wisnioski}, \& {Woods}}]{pracy16}
{Pracy}, M.~B., {Ching}, J.~H.~Y., {Sadler}, E.~M., {et~al.} 2016, \mnras, 460,
  2

\bibitem[{{Rosario} {et~al.}(2012){Rosario}, {Santini}, {Lutz}, {Shao},
  {Maiolino}, {Alexander}, {Altieri}, {Andreani}, {Aussel}, {Bauer}, {Berta},
  {Bongiovanni}, {Brandt}, {Brusa}, {Cepa}, {Cimatti}, {Cox}, {Daddi}, {Elbaz},
  {Fontana}, {F{\"o}rster Schreiber}, {Genzel}, {Grazian}, {Le Floch},
  {Magnelli}, {Mainieri}, {Netzer}, {Nordon}, {P{\'e}rez Garcia}, {Poglitsch},
  {Popesso}, {Pozzi}, {Riguccini}, {Rodighiero}, {Salvato}, {Sanchez-Portal},
  {Sturm}, {Tacconi}, {Valtchanov}, \& {Wuyts}}]{rosario12}
{Rosario}, D.~J., {Santini}, P., {Lutz}, D., {et~al.} 2012, \aap, 545, A45

\bibitem[{{Rosario} {et~al.}(2013){Rosario}, {Trakhtenbrot}, {Lutz}, {Netzer},
  {Trump}, {Silverman}, {Schramm}, {Lusso}, {Berta}, {Bongiorno}, {Brusa},
  {F{\"o}rster-Schreiber}, {Genzel}, {Lilly}, {Magnelli}, {Mainieri},
  {Maiolino}, {Merloni}, {Mignoli}, {Nordon}, {Popesso}, {Salvato}, {Santini},
  {Tacconi}, \& {Zamorani}}]{rosario13}
{Rosario}, D.~J., {Trakhtenbrot}, B., {Lutz}, D., {et~al.} 2013, \aap, 560, A72

\bibitem[{{Sadler} {et~al.}(2002){Sadler}, {Jackson}, {Cannon}, {McIntyre},
  {Murphy}, {Bland-Hawthorn}, {Bridges}, {Cole}, {Colless}, {Collins}, {Couch},
  {Dalton}, {De Propris}, {Driver}, {Efstathiou}, {Ellis}, {Frenk},
  {Glazebrook}, {Lahav}, {Lewis}, {Lumsden}, {Maddox}, {Madgwick}, {Norberg},
  {Peacock}, {Peterson}, {Sutherland}, \& {Taylor}}]{sadler02}
{Sadler}, E.~M., {Jackson}, C.~A., {Cannon}, R.~D., {et~al.} 2002, \mnras, 329,
  227

\bibitem[{{Schinnerer} {et~al.}(2010){Schinnerer}, {Sargent}, {Bondi}, {Smol{\v
  c}i{\'c}}, {Datta}, {Carilli}, {Bertoldi}, {Blain}, {Ciliegi}, {Koekemoer},
  \& {Scoville}}]{schinnerer10}
{Schinnerer}, E., {Sargent}, M.~T., {Bondi}, M., {et~al.} 2010, \apjs, 188, 384

\bibitem[{{Schinnerer} {et~al.}(2007){Schinnerer}, {Smol{\v c}i{\'c}},
  {Carilli}, {Bondi}, {Ciliegi}, {Jahnke}, {Scoville}, {Aussel}, {Bertoldi},
  {Blain}, {Impey}, {Koekemoer}, {Le Fevre}, \& {Urry}}]{schinnerer07}
{Schinnerer}, E., {Smol{\v c}i{\'c}}, V., {Carilli}, C.~L., {et~al.} 2007,
  \apjs, 172, 46

\bibitem[{{Schmidt}(1968)}]{schmidt68}
{Schmidt}, M. 1968, \apj, 151, 393

\bibitem[{{Schmidt} \& {Green}(1983)}]{sg83}
{Schmidt}, M. \& {Green}, R.~F. 1983, \apj, 269, 352

\bibitem[{{Schneider} {et~al.}(2010){Schneider}, {Richards}, {Hall}, {Strauss},
  {Anderson}, {Boroson}, {Ross}, {Shen}, {Brandt}, {Fan}, {Inada}, {Jester},
  {Knapp}, {Krawczyk}, {Thakar}, {Vanden Berk}, {Voges}, {Yanny}, {York},
  {Bahcall}, {Bizyaev}, {Blanton}, {Brewington}, {Brinkmann}, {Eisenstein},
  {Frieman}, {Fukugita}, {Gray}, {Gunn}, {Hibon}, {Ivezi{\'c}}, {Kent}, {Kron},
  {Lee}, {Lupton}, {Malanushenko}, {Malanushenko}, {Oravetz}, {Pan}, {Pier},
  {Price}, {Saxe}, {Schlegel}, {Simmons}, {Snedden}, {SubbaRao}, {Szalay}, \&
  {Weinberg}}]{schneider10}
{Schneider}, D.~P., {Richards}, G.~T., {Hall}, P.~B., {et~al.} 2010, \aj, 139,
  2360

\bibitem[{{Smith} {et~al.}(2014){Smith}, {Jarvis}, {Hardcastle}, {Vaccari},
  {Bourne}, {Dunne}, {Ibar}, {Maddox}, {Prescott}, {Vlahakis}, {Eales},
  {Maddox}, {Smith}, {Valiante}, \& {de Zotti}}]{smith14}
{Smith}, D.~J.~B., {Jarvis}, M.~J., {Hardcastle}, M.~J., {et~al.} 2014, \mnras,
  445, 2232

\bibitem[{{Smol{\v c}i{\'c}} {et~al.}(2017{\natexlab{a}}){Smol{\v c}i{\'c}},
  {Delvecchio}, {Zamorani}, {Baran}, {Novak}, {Delhaize}, {Schinnerer},
  {Berta}, {Bondi}, {Ciliegi}, {Capak}, {Civano}, {Karim}, {Le Fevre},
  {Ilbert}, {Laigle}, {Marchesi}, {McCracken}, {Tasca}, {Salvato}, \&
  {Vardoulaki}}]{smolcic17b}
{Smol{\v c}i{\'c}}, V., {Delvecchio}, I., {Zamorani}, G., {et~al.}
  2017{\natexlab{a}}, \aap, 602, A2

\bibitem[{{Smol{\v c}i{\'c}} {et~al.}(2017{\natexlab{b}}){Smol{\v c}i{\'c}},
  {Novak}, {Bondi}, {Ciliegi}, {Mooley}, {Schinnerer}, {Zamorani}, {Navarrete},
  {Bourke}, {Karim}, {Vardoulaki}, {Leslie}, {Delhaize}, {Carilli}, {Myers},
  {Baran}, {Delvecchio}, {Miettinen}, {Banfield}, {Balokovi{\'c}}, {Bertoldi},
  {Capak}, {Frail}, {Hallinan}, {Hao}, {Herrera Ruiz}, {Horesh}, {Ilbert},
  {Intema}, {Jeli{\'c}}, {Kl{\"o}ckner}, {Krpan}, {Kulkarni}, {McCracken},
  {Laigle}, {Middleberg}, {Murphy}, {Sargent}, {Scoville}, \&
  {Sheth}}]{smolcic17a}
{Smol{\v c}i{\'c}}, V., {Novak}, M., {Bondi}, M., {et~al.} 2017{\natexlab{b}},
  \aap, 602, A1

\bibitem[{{Smol{\v c}i{\'c}} {et~al.}(2017{\natexlab{c}}){Smol{\v c}i{\'c}},
  {Novak}, {Delvecchio}, {Ceraj}, {Bondi}, {Delhaize}, {Marchesi}, {Murphy},
  {Schinnerer}, {Vardoulaki}, \& {Zamorani}}]{smolcic17c}
{Smol{\v c}i{\'c}}, V., {Novak}, M., {Delvecchio}, I., {et~al.}
  2017{\natexlab{c}}, \aap, 602, A6

\bibitem[{{Stanley} {et~al.}(2017){Stanley}, {Alexander}, {Harrison},
  {Rosario}, {Wang}, {Aird}, {Bourne}, {Dunne}, {Dye}, {Eales}, {Knudsen},
  {Micha{\l}owski}, {Valiante}, {De Zotti}, {Furlanetto}, {Ivison}, {Maddox},
  \& {Smith}}]{stanley17}
{Stanley}, F., {Alexander}, D.~M., {Harrison}, C.~M., {et~al.} 2017, \mnras,
  472, 2221

\bibitem[{{Stanley} {et~al.}(2015){Stanley}, {Harrison}, {Alexander},
  {Swinbank}, {Aird}, {Del Moro}, {Hickox}, \& {Mullaney}}]{stanley15}
{Stanley}, F., {Harrison}, C.~M., {Alexander}, D.~M., {et~al.} 2015, \mnras,
  453, 591

\bibitem[{{Suh} {et~al.}(2017){Suh}, {Civano}, {Hasinger}, {Lusso}, {Lanzuisi},
  {Marchesi}, {Trakhtenbrot}, {Allevato}, {Cappelluti}, {Capak}, {Elvis},
  {Griffiths}, {Laigle}, {Lira}, {Riguccini}, {Rosario}, {Salvato},
  {Schawinski}, \& {Vignali}}]{suh17}
{Suh}, H., {Civano}, F., {Hasinger}, G., {et~al.} 2017, \apj, 841, 102

\bibitem[{{Suh} {et~al.}(2019){Suh}, {Civano}, {Hasinger}, {Lusso}, {Marchesi},
  {Schulze}, {Onodera}, {Rosario}, \& {Sanders}}]{suh19}
{Suh}, H., {Civano}, F., {Hasinger}, G., {et~al.} 2019, \apj, 872, 168

\bibitem[{{Terashima} \& {Wilson}(2003)}]{terashima03}
{Terashima}, Y. \& {Wilson}, A.~S. 2003, \apj, 583, 145

\bibitem[{{van der Wel} {et~al.}(2014){van der Wel}, {Franx}, {van Dokkum},
  {Skelton}, {Momcheva}, {Whitaker}, {Brammer}, {Bell}, {Rix}, {Wuyts},
  {Ferguson}, {Holden}, {Barro}, {Koekemoer}, {Chang}, {McGrath},
  {H{\"a}ussler}, {Dekel}, {Behroozi}, {Fumagalli}, {Leja}, {Lundgren},
  {Maseda}, {Nelson}, {Wake}, {Patel}, {Labb{\'e}}, {Faber}, {Grogin}, \&
  {Kocevski}}]{vdw14}
{van der Wel}, A., {Franx}, M., {van Dokkum}, P.~G., {et~al.} 2014, \apj, 788,
  28

\bibitem[{{Whitaker} {et~al.}(2012){Whitaker}, {van Dokkum}, {Brammer}, \&
  {Franx}}]{whitaker12}
{Whitaker}, K.~E., {van Dokkum}, P.~G., {Brammer}, G., \& {Franx}, M. 2012,
  \apjl, 754, L29

\bibitem[{{White} {et~al.}(2007){White}, {Helfand}, {Becker}, {Glikman}, \& {de
  Vries}}]{white07}
{White}, R.~L., {Helfand}, D.~J., {Becker}, R.~H., {Glikman}, E., \& {de
  Vries}, W. 2007, \apj, 654, 99

\bibitem[{{White} {et~al.}(2015){White}, {Jarvis}, {H{\"a}u{\ss}ler}, \&
  {Maddox}}]{white15}
{White}, S.~V., {Jarvis}, M.~J., {H{\"a}u{\ss}ler}, B., \& {Maddox}, N. 2015,
  \mnras, 448, 2665

\bibitem[{{White} {et~al.}(2017){White}, {Jarvis}, {Kalfountzou}, {Hardcastle},
  {Verma}, {Cao Orjales}, \& {Stevens}}]{white17}
{White}, S.~V., {Jarvis}, M.~J., {Kalfountzou}, E., {et~al.} 2017, \mnras, 468,
  217

\bibitem[{{Yun} {et~al.}(2001){Yun}, {Reddy}, \& {Condon}}]{yun01}
{Yun}, M.~S., {Reddy}, N.~A., \& {Condon}, J.~J. 2001, \apj, 554, 803

\end{thebibliography}

\end{document}